\documentclass[conference]{IEEEtran}
\usepackage{cite}
\usepackage{amsmath,amssymb,amsfonts}
\usepackage{algorithmic}
\usepackage{graphicx}
\usepackage{textcomp}
\usepackage{xcolor}
\usepackage[hyphens]{url}

\usepackage{etoolbox} % for newtoggle

\def\BibTeX{{\rm B\kern-.05em{\sc i\kern-.025em b}\kern-.08em
    T\kern-.1667em\lower.7ex\hbox{E}\kern-.125emX}}

\newcommand{\bheading}[1]{{\vspace{2pt}\noindent{\textbf{#1}}\hspace{2pt}}}
\newcommand{\refeq}[1]{Eq (\ref{#1})}
\newcommand{\reffig}[1]{Fig.~\ref{#1}}
\newcommand{\refsec}[1]{Section~\ref{#1}}
\newcommand{\reftbl}[1]{TABLE~\ref{#1}}

\newcommand{\sarr}{SA-RR}
\newcommand{\nofill}{NoFill}
\newcommand{\nofillbit}{NoFill}
\newcommand{\snofill}{NoSpecFill}

\newcommand{\rasfull}{Random and Safe}
\newcommand{\ras}{RaS}
\newcommand{\shtfull}{Safe History Buffer}
\newcommand{\sht}{SHB}
\newcommand{\shtfetch}{\sht fetch}
\newcommand{\shtnotify}{\textit{NoFillClear}}

\newcommand{\rasspec}{\ras-Spec}
\newcommand{\rasall}{\ras+}

\newcommand{\opsetup}{\emph{Setup}}
\newcommand{\opauth}{\emph{Authorize}}
\newcommand{\opaccess}{\emph{Access}}
\newcommand{\opuse}{\emph{Use}}
\newcommand{\opsend}{\emph{Send}}
\newcommand{\oprecv}{\emph{Receive}}

\newenvironment{packeditemize}{
\begin{list}{$\bullet$}{
\setlength{\labelwidth}{8pt}
\setlength{\itemsep}{0pt}
\setlength{\leftmargin}{\labelwidth}
\addtolength{\leftmargin}{\labelsep}
\setlength{\parindent}{0pt}
\setlength{\listparindent}{\parindent}
\setlength{\parsep}{0pt}
\setlength{\topsep}{3pt}}}{\end{list}}

\newtoggle{flagthesis}
\togglefalse{flagthesis}
\newcommand{\ifthesis}[1]{\iftoggle{flagthesis}{#1}{}}
\newcommand{\ifelsethesis}[2]{\iftoggle{flagthesis}{#1}{#2}}
\title{Random and Safe Cache Architecture to Defeat Cache Timing Attacks}
\author{Guangyuan Hu and Ruby B. Lee\\
        Princeton University \\
        \textit{\{gh9,rblee\}@princeton.edu}}
\begin{document}

\maketitle
\thispagestyle{plain}
\pagestyle{plain}

%%%%%% -- PAPER CONTENT STARTS-- %%%%%%%%

\begin{abstract}

Caches have been exploited to leak secret information due to the different times they take to handle memory accesses. Cache timing attacks include non-speculative cache side and covert channel attacks and cache-based speculative execution attacks. We first present a systematic view of the attack and defense space and show that no existing defense has addressed all cache timing attacks, which we do in this paper. We propose Random and Safe (RaS) cache architectures to decorrelate cache state changes from memory requests. RaS fills the cache with ``safe'' cache lines that are likely to be used in the future, rather than with demand-fetched, security-sensitive lines. RaS lifts the restriction on cache fills for accesses that become safe when speculative execution is resolved and authorized. Our RaS-Spec design against cache-based speculative execution attacks has a low 3.8\% average performance overhead. RaS+ variants against both speculative and non-speculative attacks have security-performance trade-offs ranging from 7.9\% to 45.2\% average overhead.

\end{abstract}

\section{Introduction}

The cache subsystem is one of modern computers' most important performance optimization features. Unfortunately, it is also the favorite target of attackers for leaking secret information in cache side channels and covert channel attacks. Hence, designing secure and performant caches is essential for future computers.

We define a \textit{secure cache} as one that is resilient to cache timing attacks. We include speculative execution attacks that send secret information out through a cache channel and non-speculative cache side or covert channel attacks. 

Partitioning cache resources between security domains is the most straightforward solution for a defense to isolate the cache state from an attacker. However, this requires substantial knowledge of programs running in a system to manage security domains properly and causes scalability and performance concerns. Furthermore, partitioning cannot prevent operation-based attacks caused solely by the victim's behaviors.

We provide a systematization of generalized hardware defense mechanisms and show that they cover all previously published defenses. Past work considered either the cache-based speculative execution attacks or the traditional non-speculative cache side-channel attacks, but not both. A future secure cache design should mitigate both attack families, which we do in this paper.

We propose a novel, decorrelating cache architecture that makes cache line fills independent of memory requests, thus leaking no information about the victim's usage to the attacker observing cache state.
Our key insight is that the major vulnerability of today's caches, exploited by cache timing attacks, is the \textbf{predictability} of \textit{which} memory line is brought into the cache, \textit{where} it is placed in the cache, and \textit{when} this happens. A requested memory line is brought into the cache, evicting and replacing an existing cache line in the same cache set, on a demand fetch. 
%Empirical studies have shown that this leads to future cache hits due to temporal and spatial locality, thus reducing the average memory access time. 
The success of cache timing attacks has shown that this traditional cache design is insecure.

In contrast, we propose a new paradigm for designing secure caches, completely changing traditional demand fetch and replacement policies. First, we dissociate the cache fill from the data fetch to the processor. We propose that the cache is only filled by what we call `safe' addresses and that this filling of the cache does not happen on a cache miss. Also, the cache set of the victim's memory request is not related to the cache set of an evicted or replaced cache line. Hence, we drop the deterministic demand fetch and LRU replacement policies, so any cache state an attacker can observe through a cache timing attack is decorrelated with the victim's memory accesses. We retain the set-associative cache design for its known benefits and for easier adoption.

To enable the cache to be still performant in reducing average memory access time, we strive to bring memory lines into the cache that will likely be used, which is the same goal as conventional caches. We show that many memory accesses are actually authorized quickly and introduce a new \shtnotify~feature to clear their no-fill status. This eliminates the need for a second load on authorization, allows early and safe cache fills and improves the performance. We experiment with fetching addresses generated by adding randomization to safe addresses collected dynamically at runtime. While this novel design methodology for secure caches can take many forms, we illustrate with two designs of \rasspec~and \rasall~with different security and performance trade-offs.

Our key contributions are:

\begin{packeditemize}

    \item A systematic characterization of cache timing attacks, including the operation-based attacks not thoroughly evaluated in past defenses and recent attacks on LRU replacement state (\refsec{sec_background_sht}).

    \item Identifying general defense mechanisms and analyzing existing defenses against cache timing attacks, showing what attacks each defense covers (\refsec{sec_hw_def_sht}).

    \item A new direction for designing decorrelation caches to prevent an attacker from observing a victim's memory usage and defeat all considered attacks. Our \ras~defense preserves the set-associative cache structure and does not require the knowledge and effort to define separate security domains (\refsec{sec_ras_arch}).

    \item A separate fetch mechanism to fill the cache for performance while leaking no information. We use a new \shtfull~(\sht) to collect safe addresses, enhanced with randomization techniques to thwart the attacker.
% and identifying the requirements needed to defeat reuse and contention-based side-%channel attacks. (I am not sure this goes together - may cut to save space)

    \item %Showing the insight that {\color{red} 46.48\% of speculative loads in benign programs that have a L1D cache miss} can be authorized quickly to lift the restriction on cache fills. 
    We design a performance optimization feature called \shtnotify~to minimize the number of times that the cache cannot be filled, thus significantly improving performance.

    \item Security evaluations on gem5 simulations using a suite of speculative execution attacks and cache side-channel attacks to verify the security of two different \ras~system designs. The security testing code will be made available to the computer architecture community upon publication.

    \item Performance overhead of only 3.8\% to defeat speculative execution attacks and security-performance trade-offs in defeating both speculative and non-speculative cache timing attacks.

\end{packeditemize}
\section{Background: Attacks}
\label{sec_background_sht}

We present a systematic view of cache timing attacks, which this paper aims to mitigate. The systematization allows us to 
%propose a new type of attack called the volume-reuse attack and 
analyze the security coverage provided by existing hardware defenses against these categories.

\ifthesis{
\begin{table}[t]
    \centering
    \ifelsethesis{
        \includegraphics[width=0.7\linewidth]{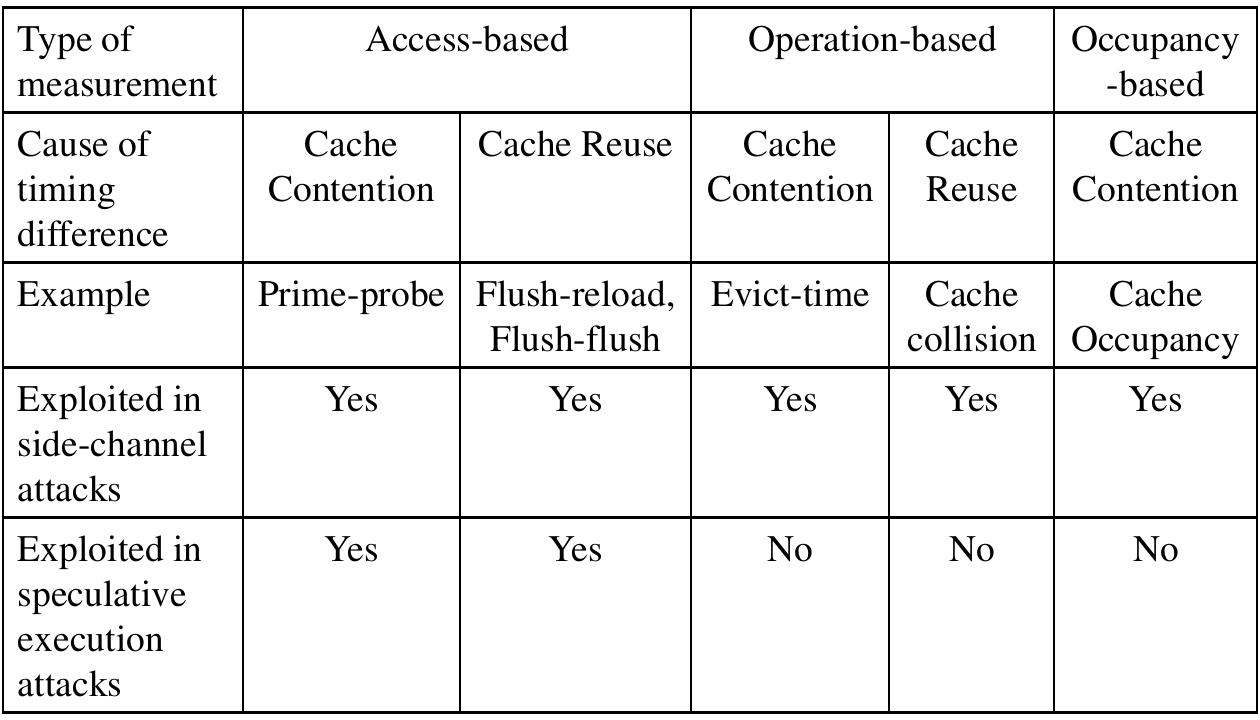}
    }
    {
        \includegraphics[width=\linewidth]{figures/side_channel_dimension.pdf}
    }
    \caption{Categories of cache side-channel attacks.}
    \label{tbl_side_channel_dimension}
\end{table}
}

\subsection{Cache Side Channels}
\label{sec_bg_side_channel_ras}

\ifthesis{Cache states affect the time taken to handle a memory request depending on whether a cache line is in the cache.} In a cache side-channel attack, a victim accesses an address determined by a secret, causing cache fills and replacements. The change in the cache state gives the attacker different timing measurements, leaking the secret information.

The difference between traditional side-channel attacks and the following cache-based speculative execution attacks is that the former does not require speculative execution or out-of-order execution features to succeed. \ifthesis{We refer to traditional side-channel attacks also as non-speculative side-channel attacks.}

The columns of \ifelsethesis{\reftbl{tbl_side_channel_dimension}}{\reftbl{tbl_matrix_defense_dimension}} show the major categories of cache channels that are exploited in cache side-channel attacks. The first dimension of cache channels is whether the attacker measures the time of his accesses, the execution time of the victim or the LRU replacement state. The second dimension is whether the attack is reuse-based or contention-based.
\ifthesis{Reuse-based attackers try to get a faster timing when accessing (reusing) the address used by the victim. Contention-based attacks observe whether the victim's access is mapped to the same set as an existing cache line, which leads to longer access time.}
The third dimension is whether the attack is speculative or non-speculative.

\bheading{Access-based attackers} \ifthesis{learn about the victim's access pattern by accessing and detecting the presence of memory addresses in the cache. Access-based attackers must co-locate on the same physical machine as the victim to perform memory accesses and measure the time taken.

Access-based attackers} typically require three steps to (1) prepare initial states, (2) let the victim execute, and (3) measure the final states with the attacker's accesses. The contention-based \textbf{prime-probe} attacker \cite{primeprobe, primeprobepercival} first accesses a set of addresses to fill the cache (priming). After the victim's execution, the attacker reaccesses the same set of addresses and observes a longer access time for cache sets that the victim evicts.

The reuse-based \textbf{flush-reload} attack \cite{flushreload2011, flushreload2014} requires shared memory with the victim. The attacker first flushes shared addresses from the cache. After the victim's execution, the attacker reaccesses the shared addresses and observes a shorter reload time for addresses the victim has accessed. 

The flush-flush attack \cite{flushflush} is also a reuse-based attack \ifelsethesis{and requires shared memory. The attacker also flushes shared addresses from the cache. Unlike the flush-reload attack, the attacker flushes the shared addresses again after the victim's execution. The }{where the} attacker tries to observe longer flush time for addresses the victim has accessed.

\bheading{Operation-based attackers} measure the victim's secret-dependent execution time 
\ifthesis{if an attacker can capture the beginning and end of the victim's execution. The execution time is longer if the victim has more secret-dependent cache misses or shorter if the victim program has more secret-dependent cache hits. Operation-based attacks can be launched remotely \cite{bernsteinattack}, which does not require the attacker's program to run on the same machine as the victim}. 
Examples of operation-based attacks are the contention-based \textbf{evict-time} attack \cite{bernsteinattack} and the reuse-based \textbf{cache collision} attack \cite{cachecollision}. 

In the evict-time attack, an eviction of a cache set happens before the victim's execution. The eviction can be caused by either the attacker or the victim itself. Certain secret values will cause the victim to use an evicted cache line, making the execution time longer.
%If the eviction is caused by the victim's accesses within its domain, the attack cannot be mitigated by partitioning-based defenses.

In a cache collision attack, the victim accesses an address and brings the cache line into the cache. Certain secret values lead to a later memory access to the same cache line, and the cache line is reused, making the execution time shorter. \ifthesis{The cache collision attack does not require the attacker to perform any accesses.} A cache collision attack is always within the victim's domain and cannot be mitigated by partitioning-based defenses.

%\bheading{Occupancy-based attackers} do similar measurements as access-based attacks but monitor the secret-dependent usage of cache lines instead of the presence of certain addresses. The cache occupancy attack \cite{cacheoccupancy} records the volume of cache contention over time by the victim to distinguish different victim behaviors.

% removed to simplify
%We further propose a new volume-reuse attack that observes the number of cache lines that the attacker can reuse after the victim's execution. A possible scenario is when the victim accesses different numbers of cache lines containing shared data depending on the secret value. An attacker can later measure the time to access all shared cache lines, and a short execution time indicates a high volume of usage by the victim. %Similar techniques can be applied to exploit shared library code.

\bheading{Attacks on LRU states} can also leak secrets \cite{dawg, lru2020hpca}. \ifthesis{The attacker preloads a few cache lines into a target cache set. }
Depending on the secret, the victim may or may not access a preloaded cache line. The victim's access can modify the LRU state even if \textit{it has a cache hit and no new cache line is brought in}. \ifthesis{When the attacker later fetches some other lines into this cache set, the LRU states decide which line is replaced and evicted.} The cache line accessed by the victim can remain in the cache as it is ``most recently used'', which can be noticed by the attacker. \ifthesis{This type of leakage can leak secrets across security domains \cite{dawg} and in speculative channels \cite{lru2020hpca}.}

% delete: shorten
\iffalse
\begin{figure}[t]
    \centering
    \includegraphics[width=\linewidth]{figures/leaked_bits.pdf}
    \caption{The bits in a memory address that can be leaked in a set-associative cache using access-contention or access-reuse cache timing attacks.}
    \label{fig_leaked_bits}
\end{figure}
\fi
%\reffig{fig_leaked_bits} shows the leakable bits in a set-associative cache. Reuse-based attacks such as the flush-reload attack can leak all bits in the address except for the bits for byte selection within a cache line but may require shared memory between the attacker and the victim. Contention-based attacks do not require shared memory but leak only the index bits to select cache sets. 
%In a set-associative cache, reuse-based attacks can leak more bits than contention-based attacks but may require shared memory between the attacker and the victim. Contention-based attacks do not require shared memory but incur more noise due to unrelated cache accesses. 

\begin{figure}[t]
    \centering
    \includegraphics[width=0.6\linewidth]{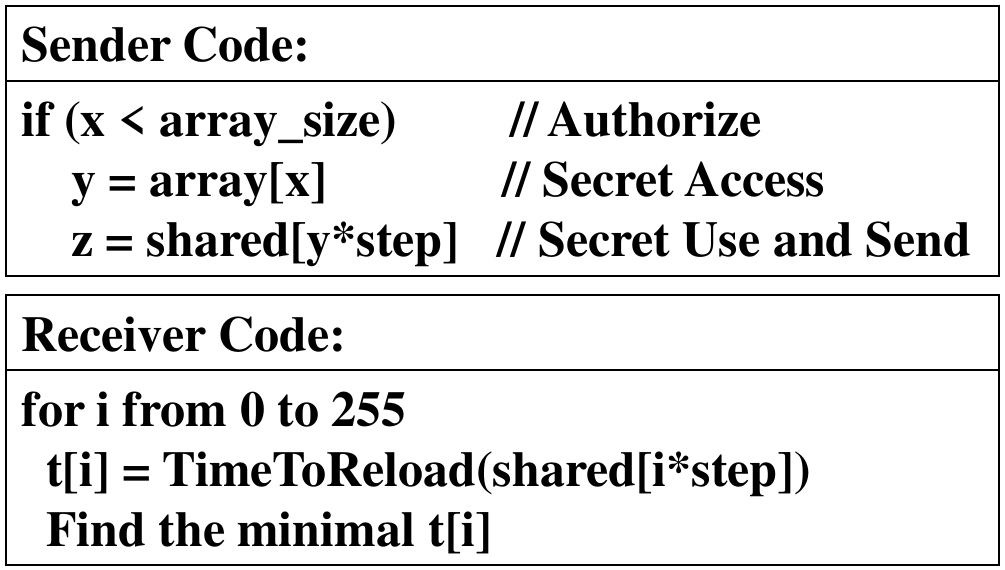}
    \caption{\normalsize{The key operations of a flush-reload Spectre-v1 Attack. 
    \ifthesis{
    Mistraining the hardware predictor and flushing all cache lines of the \textit{shared} array happen before the sender code and are not shown. Commonly used values of \textit{step}, the step size of flushing and reloading, are 64, 512 or 4096. However, the attacker can use a different value if he controls the sender.}}}
    \label{fig_code_spectre_v1}
    \vspace{-10pt}
\end{figure}

\subsection{Speculative Execution Attacks}
\label{sec_spec_attacks}

Speculative execution attacks  \cite{spectre, spectrev1112, netspectre, spectrersb, ret2spec, spectressb, spectrev3a, lazyfpIntel, meltdown, foreshadow, foreshadowNg, ridl, intelMDS, zombieload, fallout, TAA, VRS, cacheout, intelL1DES, crosstalk, intelSRBDS, lvi} are microarchitectural attacks that leak a secret even if proper permission check is enforced. The attacker exploits various hardware vulnerabilities, which have been systematically analyzed \cite{AttackSummaryMS, AttackSummary2019Graz, AttackSummary2021Yale, AttackModel2021hpca, AttackSoK2021seed}, to trigger malicious speculative execution of the sender code. The sender code transiently bypasses the access permission check, performs an illegal access to a secret and transmits it through a microarchitectural covert channel (covert sending). The secret in architectural registers will be cleared when the access is found to be illegal. However, the microarchitectural state change is not backed out of, which can be measured by the attacker (i.e., the receiver). 

Speculative execution attacks have six common and critical attack steps \cite{AttackModel2021hpca, AttackSoK2021seed}: (1) \opsetup~the hardware state to trigger speculative execution and \opsetup~the initial state of the covert channel. (2) \opauth~operation, which enforces software security but gets delayed and bypassed. (3) \opaccess~of a secret which is speculative and unauthorized. (4) \opuse~of the secret for covert sending. (5) \opsend~through a covert channel by modifying the microarchitectural state. (6) \oprecv~operation to measure the microarchitectural state and recover the secret.

\reffig{fig_code_spectre_v1} shows the attack steps in Spectre v1 attack, which mistrains the prediction for a conditional branch to transiently read a secret even if x is larger than $array\_size$. The secret is then sent through the flush-reload cache channel (\opsetup~for mistraining and preparing the cache state is not shown).

While various microarchitectural states can be used for covert sending, e.g., the contention of execution port \cite{smotherspectre}, the usage of MSHR \cite{speculativeinterference} and the branch predictor states \cite{branchscope, nda}, the cache state is still the most important unit to protect against speculative execution attacks 
\ifelsethesis{due to the following reasons:
\begin{packeditemize}
    \item There is a clear timing difference between a cache hit and a cache miss as a memory access can take a few cycles to access the L1 cache but as many as hundreds of cycles to access the main memory.
    \item Other attacks, e.g., exploiting port contention, require the attacker to launch a concurrent thread for measurement. Cache attacks do not need concurrent execution since the cache state remains after speculative execution, making the measurement easier for the attacker.
    \item Per-core states such as branch predictors can be easily cleared or invalidated at context switches. The cache contains cache lines used by different programs, making it hard to clear lines belonging to one specific thread. Also, dirty cache lines cannot be cleared with simple invalidation.
    \item While the busy state of a unit can encode 1 bit of a secret, the cache can leak multiple bits at one time if it leaks an address.
\end{packeditemize}
}
{because the cache state (1) has a clear timing difference (2) can be measured after the execution (3) has high complexity to clear.
}

The speculative execution attacks so far have been exploiting the access-based channel in our characterization of cache attacks. The operation-based channel has not been defined for speculative execution attacks. 
% deleted: volume
%\textbf{We identify that the speculative execution attacks may also leverage the volume-based channel to send the secret.}
% deleted: unclear
%As the time of squashed execution does not depend on the address of speculative memory accesses, we do not consider the operation-based channel exploitable in the last row of \reftbl{tbl_side_channel_dimension}.

%\textbf{A key difference between the speculative execution attack and the non-speculative side-channel attack is that the former's sender code can be controlled by the attacker himself, meaning the sender and the receiver are both in the attacker's domain.} In other words, the attacker can create a gadget for the secret access and sending, and run it in speculative execution. Such self-initiated same-domain attacks to bypass permission checks, such as sandboxing \cite{ret2spec} and kernel protection \cite{meltdownRepoIAIK}, are practical.

\ifelsethesis{Following the above analysis, we further distinguish non-speculative side-channel attacks (NS) and speculative execution attacks (S). In \reftbl{tbl_matrix_defense_dimension}, we consider access-based attacks in NS and S forms and the non-speculative operation-based attacks.
}
{
%Following the above analysis, we further distinguish non-speculative side-channel attacks (NS) and speculative execution attacks (S) in \reftbl{tbl_matrix_defense_dimension}.
}

\begin{table*}[t]
    \centering
    \includegraphics[width=\linewidth]{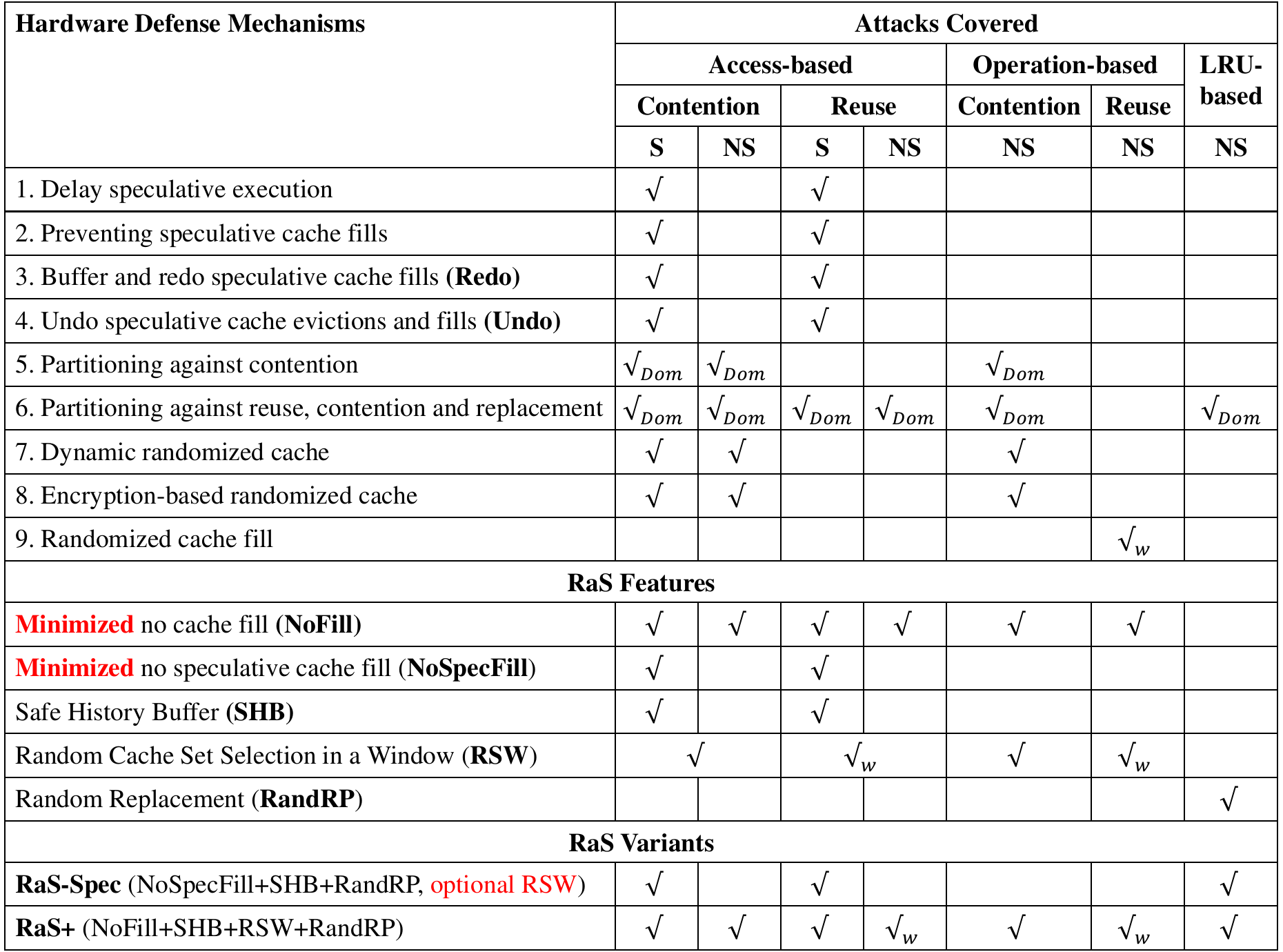}
    \caption{\normalsize{Cache timing channels: Attacks and Defense Strategies.} 
    %A (\checkmark) in the table means a defense did not claim to mitigate a specific attack, e.g., because the defense was proposed before the attack, but is considered in our analysis as being able to defeat the attack according to the design described. 
    %Details about note \{1\} are explained in \refsec{sec_hw_def_sht}. The security of \{2\} is explained in \refsec{sec_ras_arch}.
    }
    \label{tbl_matrix_defense_dimension}
    \vspace{-20pt}
\end{table*}

%\footnotetext{aa}

\section{Insights on Cache Defenses}
\label{sec_hw_def_sht}

We identify general cache defense mechanisms and systematically analyze the cache attacks each mechanism can defeat. The analysis is summarized in \reftbl{tbl_matrix_defense_dimension}.
%, which shows that existing defenses address different aspects of the attack space, but a comprehensive solution is needed.
%Some of these attacks are not addressed in the research papers using the defense mechanism, but our analysis may also credit the defense with covering these attacks. 
Near the bottom of \reftbl{tbl_matrix_defense_dimension}, we show some new defense mechanisms that we propose for our \rasfull~(\ras) architecture.

\subsection{Defenses against Speculative Execution Attacks}
\ifthesis{
Defenses can also prevent mistraining, which triggers malicious speculative execution. Predictor states can be cleared or encrypted \cite{samsung:pred:encryp, csf, ibpb:amd} against cross-domain attackers. These techniques do not prevent self-initiated speculative execution attacks or side-channel attacks that do not require mistraining.
}
%The critical attack steps that happen during a speculative execution attack are \textit{Access}, \textit{Use} and \textit{Send} (see \refsec{sec_spec_attacks}). A defense can defeat speculative execution attacks if it can prevent one of these steps. 
In \reftbl{tbl_matrix_defense_dimension}, we identify general defense approaches as being delay-based (Rows 1), prevention-based (Row 2), redo-based (Row 3) and undo-based (Row 4). %This paper addresses the covert sending phase through a cache. However, we also include Rows 1 and 2, which cover the leakage through other covert channels.

%Delay-based mechanisms can defeat speculative execution attacks. Defenses on Rows 1 and 2 can also prevent other speculative microarchitectural channels than cache-based channels. 

Delay-based defenses can delay the \opaccess~to the secret, the \opuse~of the secret or speculative cache state changes by a speculative load until the execution is authorized.
Context-sensitive Fencing \cite{csf} and secure bounds check \cite{sabc} delay the \opaccess~to a secret before the execution is verified. NDA\cite{nda}, STT \cite{stt}, SpectreGuard \cite{spectreguard}, ConTExT \cite{context} and SpecShield \cite{specshield} allow the speculative secret access but delay the use of the secret by later instructions.
%However, they only protect the secret from certain hardware units, e.g., memory and special registers.
``Delay-on-miss'' defenses delay the covert sending signal, e.g., Conditional Speculation \cite{condspec}, Efficient Speculation \cite{efficientspec} and DOLMA \cite{dolma}.

%Delay-based defenses (Rows 1-3) do not prevent non-speculative side-channel attacks where the victim has the permission to access the secret, and the victim's code will be executed.

Preventing speculative \opsend~from modifying microarchitecture states can also defeat speculative execution attacks. CFENCE \cite{csf} is a special fence operation that can be inserted before speculative loads to prevent it from filling the cache.

Redo-based defenses, e.g., InvisiSpec \cite{invisispec}, SafeSpec \cite{safespec}, MuonTrap \cite{muontrap} and GhostMinion \cite{ghostminion}, insert speculatively used cache lines into a new buffer. State changes are made to the cache once the execution is verified.

Undo-based defenses, e.g., CleanupSpec \cite{cleanupspec}, allow speculative cache state changes and restore the old state if the speculative execution is squashed. However, recent attacks \cite{unxpec} show that the undo operation can be exploited for leakage. %CleanupSpec restores old cache lines evicted from the L1 cache while avoiding restoring lines in the CEASER-type randomized LLC. CleanupSpec only protects the L1 cache against speculative volume-contention attacks but not LLC, as the attacker may still see a secret-dependent number of unrestored evictions in the LLC. 
%In side-channel attacks, the victim's execution will not be squashed, so undo-based defenses do not restore the cache state, hence not preventing the attacks.

Defenses against speculative execution attacks prevent the leak of a secret in malicious speculative execution. However, they cannot mitigate non-speculative side-channel attacks as the victim's cache state changes will finally be made.

\subsection{Defenses against Side-channel Attacks}

The main hardware defense strategies proposed to defeat side-channel attacks are cache partitioning (Rows 5-6) and cache randomization (Rows 7-9).
%Parenthetical numbers are notes explained in the text, while subscripts are conditions for the attack to be defeated.
%Subscripts are explained in the text.

\bheading{Cache partitioning.}
\ifelsethesis{Partitioning the cache resources, e.g., PLCache\cite{plrpcache}, NoMoCache\cite{nomocache}, CATalyst\cite{catalyst}, MI6\cite{mi6}, IRONHIDE\cite{ironhide}, can prevent the contention between different domains
, defeating access-based %and volume-based 
attacks. More advanced partitioning, e.g., DAWG \cite{dawg}, has also been proposed to prevent cross-domain hits, which further mitigates access-based %and volume-based 
cache reuse attacks. However, these defenses} 
{
Partitioning the cache resources can prevent contention-based attacks (Row 5: PLCache\cite{plrpcache}, NoMoCache\cite{nomocache}, CATalyst\cite{catalyst}, MI6\cite{mi6}, IRONHIDE\cite{ironhide}, HYBCACHE \cite{hybcache}, Cachelets\cite{cachelet}) between different domains (denoted $\checkmark_{dom}$ in \reftbl{tbl_matrix_defense_dimension}). Advanced partitioning, e.g., DAWG \cite{dawg} (Row 6), further prevents cross-domain cache reuse and LRU state changes. They}
do not prevent cache attacks inside the same domain. Partitioning-based defenses require the software to accurately identify security domains\ifthesis{ and even the knowledge of the hardware system to enforce way-based or set-based partition}.

\bheading{Randomized caches.} (Row 7) Defenses achieve randomized address-to-cache mappings and perform random cache line evictions against contention-based attacks, eliminating the information leak. Designs have been proposed for L1 caches, e.g., NewCache (\cite{newcache, newcache2016micro}), and for last-level caches (LLCs), e.g., MIRAGE \cite{mirage} and PhantomCache \cite{phantomcache}.
%, to achieve randomized address-to-cache mappings. 
%In same-domain speculative execution attacks, the attacker can bypass the protection and introduce deterministic contention, e.g., the index hit in NewCache \cite{newcache2013hasp}.

(Row 8) Encrypted-address caches have been proposed to achieve similar randomized address-to-cache mappings using computation-based techniques, e.g., CEASER \cite{ceaser}, CEASER-S \cite{evictionset2019} and ScatterCache \cite{scattercache}. 
%while avoiding storing mapping tables
Contention-based attackers need to first spend time creating an eviction set (\cite{llcfangfei, evictionset2019, fixit2021sp, primepruneprobe}), which is a set of addresses to evict a target cache line. Periodic remapping of cache lines is required to prevent attackers from finding an eviction set. The remapping frequency required is high for small caches, making this protection applicable only to large caches like LLCs.

%Both types of randomized caches do not address reuse-based cache timing attacks. %Also, they randomize the location of cache contention but cannot hide the volume of cache usage against volume-based attacks.

\bheading{Randomized cache fill.}
(Row 9) The Random Fill cache \cite{randomfill} is a randomization-based secure cache that changes the fetch-and-fill policy. Instead of filling the exact requested address (\textbf{demand fetch}) into the cache, the Random Fill cache randomly fetches cache lines within the neighborhood of the requested memory line. If this neighborhood covers the whole memory region referred to by a secret-dependent address (denoted $\checkmark_w$ in \reftbl{tbl_matrix_defense_dimension}), e.g., a lookup table for encryption, Random Fill cache defeats cache collision attacks.

However, fetching adjacent addresses of a speculative load can still leak information in speculative execution attacks. Furthermore, the original Random Fill cache only prevents security-sensitive cache fills in the L1 cache, leaving the other cache levels unprotected.
%However, in the same-domain speculative execution attacks, the attacker controls the sender code and may choose an arbitrarily large region to bypass the defense. Random Fill cache does not prevent volume-based attacks as the frequency of creating random fetches depends on the demand fetches. Random Fill cache is not designed to mitigate contention-based attacks either.

\ifthesis{
    \begin{figure*}[t]
        \centering
        \includegraphics[width=\linewidth]{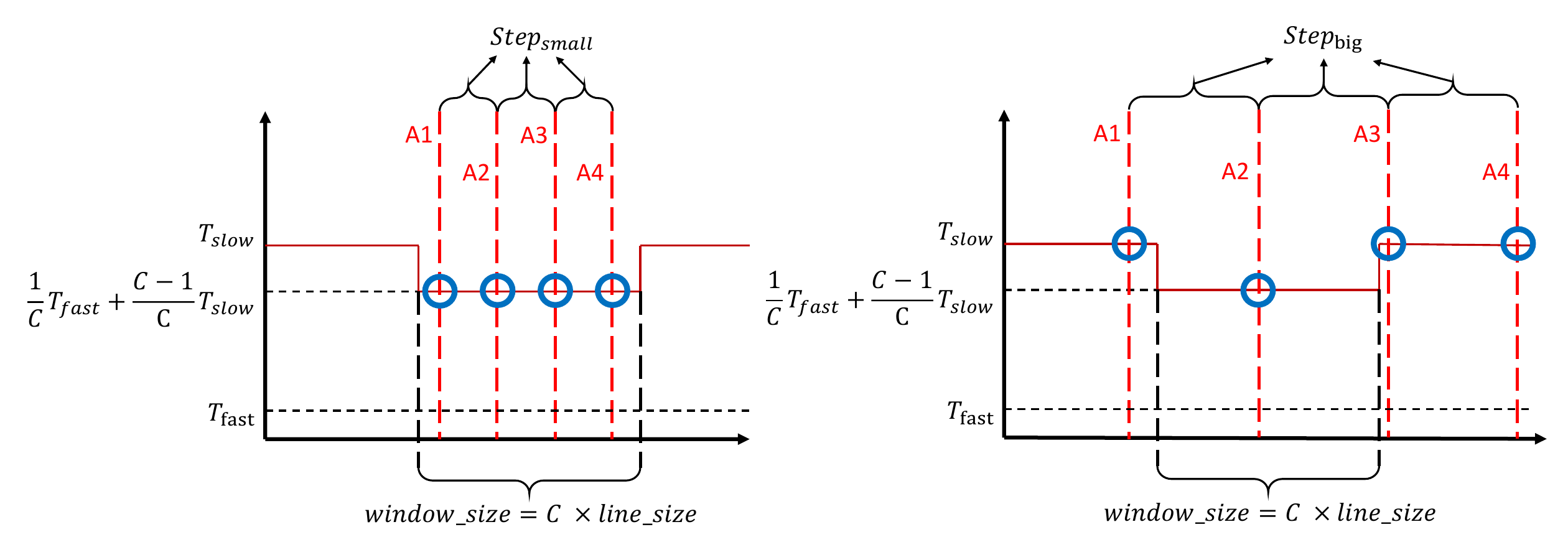}
        \caption{An example to leak a 2-bit secret of \textit{01} by flushing and reloading $2^2 =4$ addresses. Random fetching within a window can prevent reuse-based cache timing attacks if the window size is large enough so that all possible reload addresses (A1 to A4) are in the window (left). However, suppose the step size is relatively large, and the window does not cover all addresses. In that case, the average latency (circles in the figure) to access different addresses will be different, causing the secret to be leaked (right).}
        \label{fig_neighborhood_flush_reload}
    \end{figure*}

    We can show that the Random Fill cache cannot defeat all cache-based speculative execution attacks. In the following example in \reffig{fig_neighborhood_flush_reload}, we consider the attacker-controlled step size (in line 3 of \reffig{fig_code_spectre_v1}) for the Spectre v1 attack, The left part of \reffig{fig_neighborhood_flush_reload} shows the effectiveness of window-based random fetching against the attack. If a window containing $C$ cache lines is used, each cache line in the window has a chance of $\frac{1}{C}$ to be fetched, leading to a short access time of $T_{fast}$, and a chance of $\frac{C-1}{C}$ for not being fetched, leading to a longer access time of $T_{slow}$. If the measurement is performed many times by the attacker, all the addresses in the window will have an average access time of $\frac{1}{C}T_{fast} + \frac{C-1}{C}T_{slow}$. If the measurable addresses, A1 to A4, are all within the window, the attacker cannot observe any timing difference and hence gets defeated.
    
    However, if the step size between secret-dependent addresses is big so that the window cannot cover all of these addresses, the security against reuse-based attacks is compromised. In the right part of \reffig{fig_neighborhood_flush_reload}, we show a memory access pattern that is not protected. The 2-bit secret value of \textit{01} is supposed to cause a demand fetch of the address A2. The random fill architecture does not directly fill this address into the cache. Instead, it makes a random fetch in the neighborhood, causing an average latency of $\frac{1}{C}T_{fast} + \frac{C-1}{C}T_{slow}$ to access A2 later. With the big step size, other sensitive addresses, i.e., A1, A3 and A4, do not fall into this window and have the access latency of $T_{slow}$. The 2-bit secret can be revealed to be 01. While the timing difference between $\frac{1}{C}T_{fast} + \frac{C-1}{C}T_{slow}$ for reloading A2 and $T_{slow}$ for reloading A1, A3 and A4 can be small, an intelligent attacker can measure multiple adjacent cache lines near each reload address to aggregate the reload time and observe more evident timing difference.
    
    It is shown that the security of randomly fetching addresses in a window depends on whether the window size is large enough. For a cryptographic algorithm to defend itself against cache side-channel attacks, the victim knows the size of lookup tables and can choose a proper window size for protection. However, in the self-initiated speculative execution attacks, the attacker controls the sender code and can select an arbitrarily large step, e.g., the variable \textit{step} in \reffig{fig_code_spectre_v1}, to bypass the defense.
    
    We take some inspiration from the Random Fill cache and add the notion of safe addresses, which we could collect in a \shtfull~and use to fill the cache to design our \rasfull~cache. While Random Fill cache only defeats cache collision side channels, our \ras~cache can also defeat speculative attacks and non-speculative prime-probe and flush-reload attacks.

    %Therefore, to provide absolute security against speculative execution attacks, the window for randomly fetching around the demand fetch address should cover the whole address space of a program, which will fetch many unused addresses and lead to a big performance overhead.
    
    \iffalse
    We show that \textbf{there are two aspects to guarantee security to fetch addresses in a window randomly.} The first aspect is to include all security-critical addresses in the window. The second aspect is to make the timings of $\frac{1}{C}T_{fast} + \frac{C-1}{C}T_{slow}$ for the demand fetch and $T_{slow}$ for other security-critical addresses indistinguishable, i.e., making the $T_{diff}$ in the following equation small enough. 
    \begin{align}
        T_{diff} &= T_{slow} - (\frac{1}{C}T_{fast} + \frac{C-1}{C}T_{slow}) \\
                 &= \frac{1}{C} (T_{slow} - T_{fast})
    \end{align}
    For example, in a system with a cache line size of 64 bytes, $T_{fast}$ being 2 cycles and $T_{slow}$ being 150 cycles, making $T_{diff}$ smaller than 2 cycles requires a window size $C$ of 74 cache lines (), making $T_{diff}$ smaller than 1 cycles requires a window size $C$ of 148 cache lines.
    
    Both aspects require a large window size, i.e., a large $C$.
    \fi
}

\bheading{The takeaway is that none of the proposed defenses has addressed all types of attacks.}
\begin{figure*}[t]
    \centering
    \ifelsethesis{
        \includegraphics[width=\linewidth]{figures_ras/block_diagram_full.pdf}
    }
    {
        \includegraphics[width=0.86\linewidth]{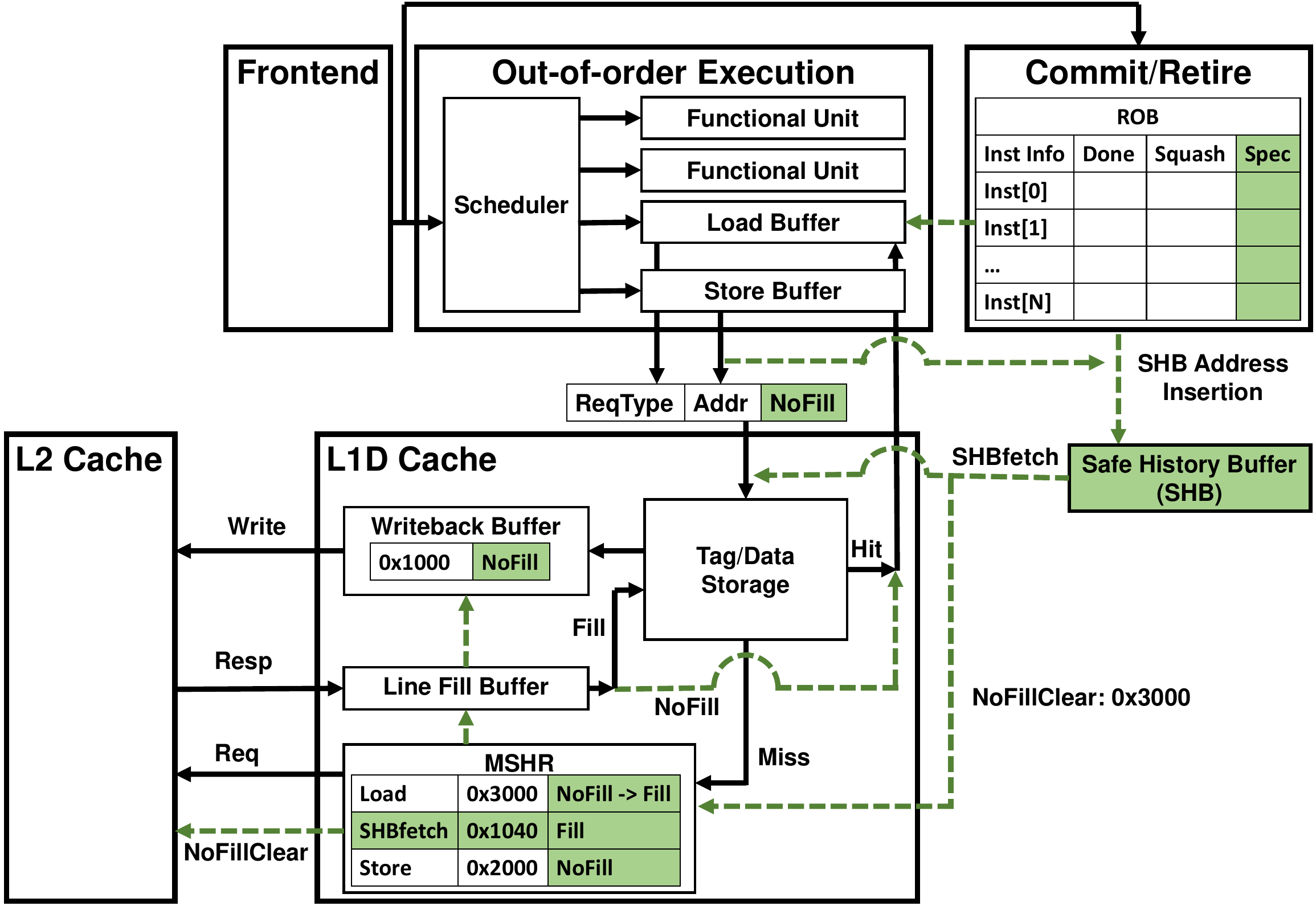}
    }
    \caption{\normalsize{The block diagram of Random and Safe (\ras) cache architecture. New units, signals and fields are shown in green.}}
    \label{fig_block_diagram}
    \vspace{-10pt}
\end{figure*}

\section{\ras~Architecture}
\label{sec_ras_arch}

\subsection{Threat Model}
\label{sec_ras_threat_model}

We aim to defeat both speculative execution attacks exploiting the cache channel and traditional cache side-channel attacks, which are categorized as columns in \reftbl{tbl_matrix_defense_dimension}. We consider speculative execution attacks caused by branch misprediction (e.g., Spectre v1\cite{spectre}), memory disambiguation mispredicted to skip store-to-load forwarding (e.g., Spectre SSB\cite{spectressb}) and faults of previous instructions (e.g., Meltdown\cite{meltdown}). A memory instruction becomes authorized (\textbf{Safe}) when all previous instructions have finished execution without a fault. For side-channel attacks, we consider access-based and operation-based attacks that are either reuse-based or contention-based. We also prevent leakage through the LRU replacement state. We do not mitigate the leakage due to software bugs allowing an attacker to access a secret legally.

We aim to build a valid defense that does not require the system software to allocate security domains for programs, which is hard without detailed knowledge and raises scalability concerns. We assume the system software is trusted for protecting registers, if any, which enables the \ras~defense or configures \ras~parameters.

\subsection{Architecture Overview}

We propose \rasfull~(\ras) architecture where observations of the cache state by an attacker give no meaningful information about the addresses used by the program (executed by a victim). \ras~conceptually differs from previous cache architectures, which do not completely eliminate predictable cache state changes against both reuse-based and contention-based attackers. \ras~uses a combination of randomization techniques and constant rate techniques. In randomization techniques, every possible outcome is equally likely, giving maximum entropy and minimum information leakage.

Our key insights are:

(1) Filling the cache with deterministic cache line insertions and evictions is the primary source of information leaks in cache timing attacks.

(2) Decorrelating cache fills from addresses used by the program can prevent these information leaks.

(3) %The race condition between the software authorization and execution of memory instructions, which is the root cause of speculative execution attacks \cite{AttackModel2021hpca}, gives opportunities to optimize the performance while maintaining security. 
In benign programs, many speculative memory accesses are authorized before the requested cache line is returned. The restriction on cache fills by these accesses should be quickly lifted for better performance (\refsec{sec_ras_shtnotify}).

An essential contribution of this paper is to show how \ras~defenses improve security while maintaining the original goal of a cache to achieve short memory access time.

\reffig{fig_block_diagram} is a block diagram of \ras~which highlights a new field called the ``\nofillbit'' attached to memory requests to decide whether cache fills will be allowed (\refsec{sec_ras_nofill}) and a new hardware unit, called the \shtfull~for sending safe prefetch-type requests (\refsec{sec_arch_sht}). These new \ras~features are shown in green. 

%We use this two-level cache system to explain and evaluate our idea. The \ras~architecture is a general method with multiple places for the designer to choose and test various configurations to achieve the desired security-performance trade-off.

We show two prototypes of the architecture (\refsec{sec_ras_variants}). \rasspec~can defeat all speculative execution attacks with better average performance than previous speculative defenses. 
%The simplest \rasspec~achieves better average performance than previous speculative defenses with the same threat model. \rasspec~can further improve the performance by leveraging random fetching. 
Another version, \rasall~defeats both cache-based speculative execution attacks and non-speculative cache side-channel attacks. Their security scopes are summarized in the bottom rows in \reftbl{tbl_matrix_defense_dimension}.
In \refsec{sec_ras_shtnotify}, we introduce a performance optimization feature to allow secure cache fills.

The idea of \ras, i.e., disallowing speculative fills and performing randomized fetches based on safe addresses, can be extended to other storage units like the instruction cache and the translation lookaside buffer (TLB).

\ifthesis{
\begin{table*}[t]
    \centering
    \includegraphics[width=0.95\linewidth]{figures_thesis/matrix_defense_mechanism_ras.pdf}
    \caption{\normalsize{Cache timing channels addressed by \ras.} 
    %A (\checkmark) in the table means a defense did not claim to mitigate a specific attack, e.g., because the defense was proposed before the attack, but is considered in our analysis as being able to defeat the attack according to the design described. 
    %Details about note \{1\} are explained in \refsec{sec_hw_def_sht}. The security of \{2\} is explained in \refsec{sec_ras_arch}.
    }
    \label{tbl_matrix_defense_dimension_ras}
\end{table*}
}

\subsection{Cross-level \nofillbit~Chain and Protection for Writeback}
\label{sec_ras_nofill}

We mark security-sensitive memory requests with a \nofillbit~bit, propagating from the request through the miss status holding register (MSHR) entry to the Line Fill Buffer and the next-level caches. 
%In \ras, a cache hit is handled as usual for a memory request with no changes.

First, we clarify MSHR and Line Fill Buffers. 
An MSHR entry in the cache records missing memory requests to the same cache line. 
%The \nofillbit~bit is also copied into the MSHR entry. 
When a memory request gets the requested cache line from the next level of the cache-memory hierarchy, the cache line is placed in a buffer called the line fill buffer, and the \nofillbit~bit of the corresponding MSHR is checked. For a cache miss, if the \nofillbit~bit is 0, the cache miss is handled as usual: the requested line is brought (i.e., filled) into the data and tag storage of cache (the \textit{Fill} path in \reffig{fig_block_diagram}) and sent to the processor. If the \nofillbit~field is set to 1, the requested data is sent to the processor (the \textit{NoFill} path in \reffig{fig_block_diagram}), but the cache line will not fill.

We prevent new writeback attacks by propagating our \nofillbit~bit across all cache levels. For instance, if the processor performs a store to a secret-dependent address, which is marked no-fill for security. The store data is written to the Line Fill Buffer for this no-fill store. The cache line does not fill the L1D cache but is written back to the L2 cache. If the L2 cache allocates a cache line for this write-back, this cache line will be detectable as an L2 hit by the attacker. We prevent attacks by attaching a \nofillbit~bit to write-backs so that caches will directly forward the no-fill write-back to its Writeback Buffer. 

\subsection{\shtfull}
\label{sec_arch_sht}

Disallowing cache fills prevents future cache hits on these lines, causing significant performance degradation. We show how an independent cache fill mechanism can be designed to fill the cache with safe and hopefully useful lines that will result in future cache hits. We define a safe address as a non-speculative load or store address. A speculative memory address becomes safe when all previous instructions have finished execution without a fault. This can happen before an instruction commits. We introduce the \shtfull~(\sht), which collects safe addresses during the program's execution and generates cache fills to enable cache hits and improve performance while maintaining cache security. 

Two key operations are associated with an \sht: how it is populated with safe addresses and how it is used to generate a memory request to fill a line into the cache. 

\bheading{Inserting Safe Addresses into the \shtfull.} 
For speculative attacks, we define a \textit{safe address} as a memory address that is no longer speculative, that is, it will be executed and committed. We say that this memory address is \textbf{authorized}. Only authorized (safe) addresses of memory accesses can be inserted into \sht. 

%As a quick review of examples of speculation vulnerabilities exploited in different speculative attacks, a load instruction is speculative if it is in one of the paths of a conditional branch instruction and the conditional branch is not yet resolved (e.g., Spectre v1). Also, a load is speculative until all previous instructions have finished execution without a fault. Also, any store-load address disambiguation has been completed (e.g., Spectre SSB). When a memory load is no longer speculative (\textbf{Authorized}), the load cannot be used as a covert sending signal to leak out a secret. Hence, it is a safe address to insert into the \sht.

A memory store is always sent to the cache after the store instruction commits, so we insert store addresses into \sht~when the store request enters the cache.
To identify authorized memory loads, the re-order buffer (ROB), which tracks the instructions in the pipeline, is modified to mark each instruction as speculative or not. When a memory access is authorized, its address can be inserted into the \sht. 

\begin{figure}[t]
    \centering
    \ifelsethesis{
        \includegraphics[width=0.7\linewidth]{figures_ras/sht_queue_rand.pdf}
    }
    {
        \includegraphics[width=0.95\linewidth]{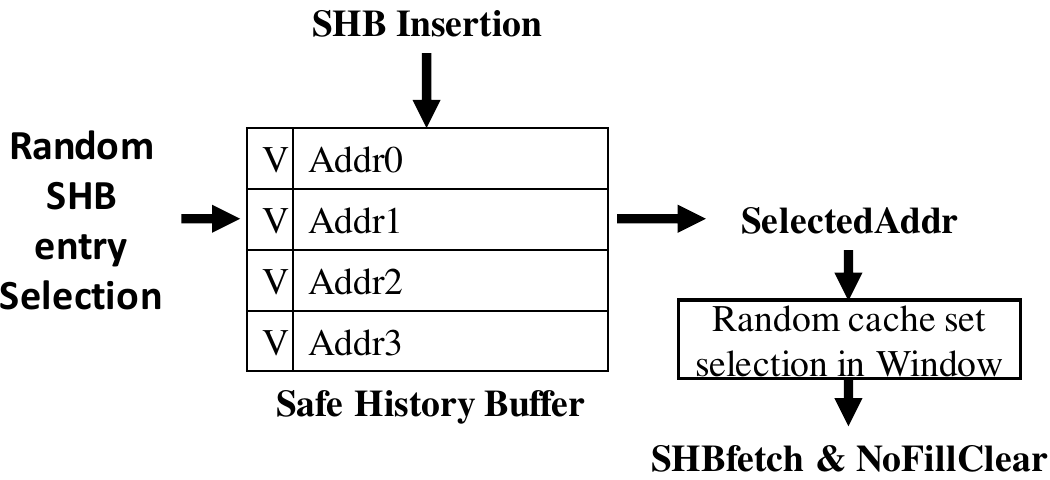}
    }
    \caption{A \shtfull~with FIFO-like \sht~insertion and random \sht~entry selection for \shtfetch. A random cache set (and associated memory line) within the \ras~Window that includes the randomly selected \sht~address is selected and sent for \shtfetch~and filled into the cache.}
    \vspace{-10pt}
    \label{fig_sht_queue_rand}
\end{figure}

\bheading{Using the Safe History Buffer.} 
This is labelled \shtfetch~in \reffig{fig_block_diagram}. We describe \textbf{\textit{what}} lines we select to fill the cache, \textbf{\textit{when}} we fill the cache, and \textbf{\textit{where}} we put the lines in the cache.

%\begin{packeditemize}
    \textbf{(1) Random Line in Window for \shtfetch.}
    While using authorized addresses for \shtfetch~prevents speculative execution attacks, fetching the original authorized address is vulnerable to side-channel attacks. In a non-speculative side-channel attack, all addresses in the victim program will become \sht~entries. Issuing \shtfetch~to the same addresses to fill into the cache
    %is equivalent to replaying the accesses and 
    can still leak the secret.
    
    \reffig{fig_sht_queue_rand} shows one possible \sht~implementation that we use in our prototype systems. First, if there are multiple \sht~entries, we randomly select an \sht~entry. Then, we select a random memory line from a window of memory lines that include the selected \sht~address. Our prototypes define the window of W lines as aligned to the W-line memory region and including the selected \sht~address. Filling a different cache line than the demand fetch address decorrelates both cache reuses and evictions in side-channel attacks.
    %The required range (window) of randomly selected lines is discussed in \ifelsethesis{\refsec{sec_analysis_window_size}}{\refsec{sec_ras_variants}}.
    
    When an \shtfetch~enters the cache, it first checks if the address is already in the \textit{Tag/Data Storage} (see \reffig{fig_block_diagram}). If the address is not found, the \shtfetch~request is inserted into MSHR as a new memory request.
    %\shtfetch~always has the \nofillbit~bit cleared.
    %If the \shtfetch~address is the same as another entry in the MSHR, it need not be a new entry in the MSHR. It changes the \nofillbit~bit of that entry with the same address to 0. This example is shown at the bottom of \reffig{fig_block_diagram}.

    \textbf{(2) Constant-rate \shtfetch.}
    One of the more unusual aspects of our \ras~architecture is how we decorrelate when cache fills happen, from when cache misses happen.
    We propose to issue \shtfetch~at a constant rate. When all the demand fetches are made no-fill, and the cache can install new lines only for the constant-rate \shtfetch, the number of victim's accesses is decoupled from cache evictions.
    %, thwarting cache occupancy attacks \cite{cacheoccupancy} that counts the evictions.
    %constant-rate \shtfetch es lead to a constant eviction rate, hence defeating the cache occupancy attack. In fact, as long as the issue rate does not depend on data used in execution, it can prevent cache occupancy attacks.
    
\bheading{Random Replacement.}
Commonly used cache replacement policies within a cache set, such as LRU and PLRU, are known to leak secrets (see \ifelsethesis{\refsec{sec_past_attack_side_channel}}{\refsec{sec_bg_side_channel_ras}}). We choose the random replacement policy to prevent such leakage. Random replacement policy is stateless, making it impossible for the attacker to record or recover secret-dependent state changes in non-speculative or speculative channels.

%\end{packeditemize}

\iffalse
\begin{figure*}[t]
    \centering
    \ifelsethesis{
        \includegraphics[width=\linewidth]{figures_ras/eval_spectre_v1_fr.pdf}
    }
    {
        \includegraphics[width=\linewidth]{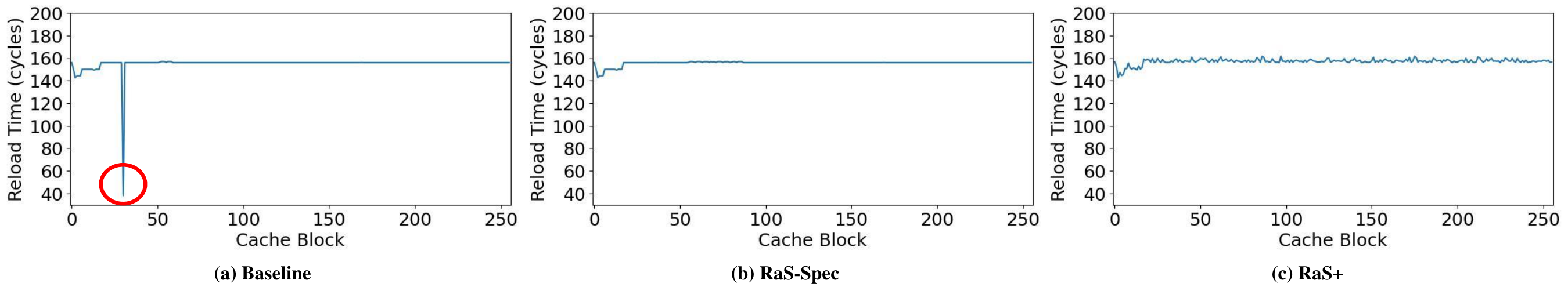}
    }
    \caption{\normalsize{The attacker's measurement in a flush-reload Spectre v1 attack. The secret value is 30.}}
    \label{fig_eval_spectre_v1_fr}
\end{figure*}

\begin{figure*}[t]
    \centering
    \ifelsethesis{
        \includegraphics[width=\linewidth]{figures_ras/eval_spectre_v1_pp.pdf}
    }
    {
        \includegraphics[width=\linewidth]{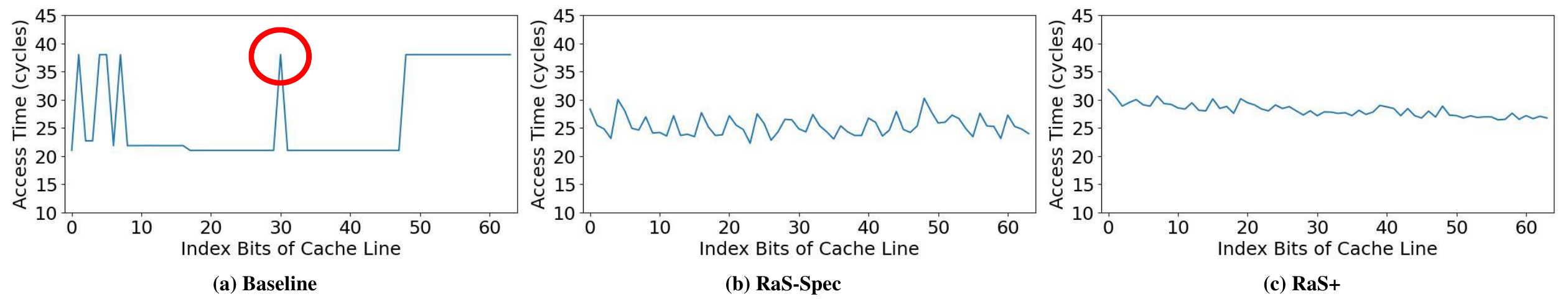}
    }
    \caption{\normalsize{The attacker's measurement in a prime-probe Spectre v1 attack. The secret value is 30.}}
    \label{fig_eval_spectre_v1_pp}
\end{figure*}
\fi

\subsection{\ras~Prototypes: RASspec and RAS+}
\label{sec_ras_variants}

We implement two \ras~prototypes, \rasspec~and \rasall. \rasspec~mitigates cache-based speculative execution attacks and, \rasall~mitigates both speculative and non-speculative cache timing attacks. Their features and security are summarized in \ifelsethesis{\reftbl{tbl_matrix_defense_dimension_ras}}{\reftbl{tbl_matrix_defense_dimension}}. 

\rasspec~implements No speculative fill (\snofill~in \ifelsethesis{\reftbl{tbl_matrix_defense_dimension_ras}}{\reftbl{tbl_matrix_defense_dimension}}): stores and non-speculative loads can fill the cache while speculative loads cannot. %Addresses of authorized memory requests are inserted into \sht~from the ROB.
RaS-Spec with a window size of 1, i.e., adding no offset, fetches the same address as the earlier no-fill speculative access after the access is authorized.
While \rasspec~already defeats speculative cache attacks without further randomization, we show that random fetching in a small window can improve the performance (see \refsec{sec_perf_rasspec}).

\rasall~implements the stronger protection of No cache fill (\nofill~in \ifelsethesis{\reftbl{tbl_matrix_defense_dimension_ras}}{\reftbl{tbl_matrix_defense_dimension}}) to mark all loads and stores as no-fill and prevent them from filling the cache.

\begin{figure}[t]
    \centering
    \ifelsethesis{
        \includegraphics[width=0.8\linewidth]{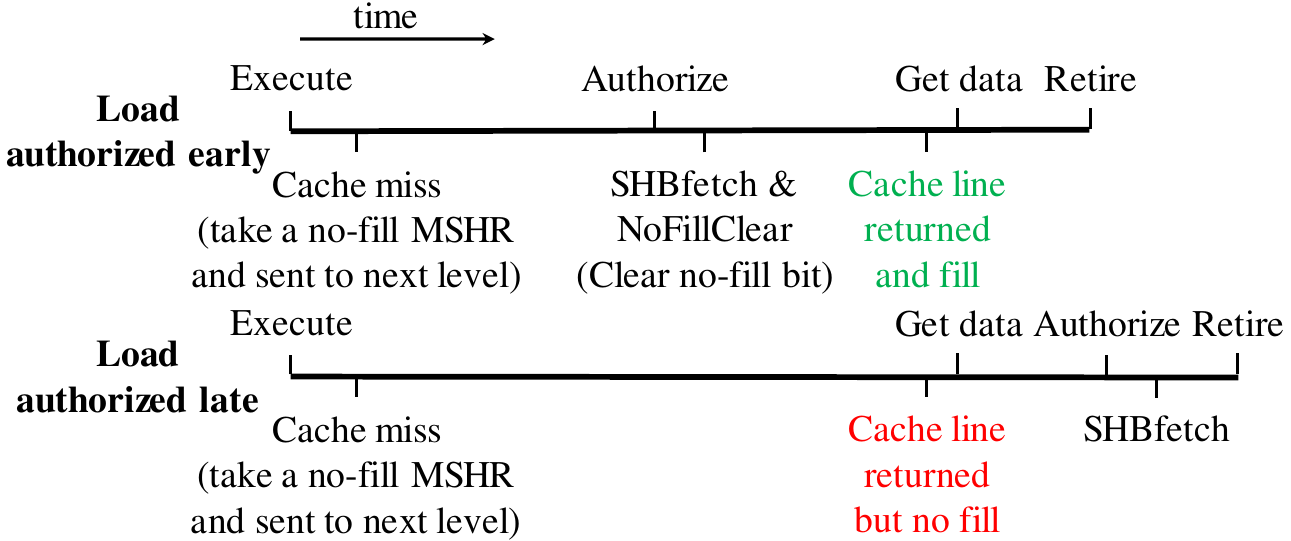}
    }
    {
        \includegraphics[width=\linewidth]{figures/timeline_authorize.pdf}
    }
    \caption{Timeline of speculative loads that are authorized before or after the cache line is returned. Minimization of no-fill MSHRs with the \shtnotify~signal enables cache line fills for future cache hits.}
    \label{fig_timeline_authorize}
    \vspace{-10pt}
\end{figure}

\subsection{Minimizing \nofill~and \snofill}
\label{sec_ras_shtnotify}

We introduce a performance optimization feature called \shtnotify~which sends the \shtfetch~address directly to MSHRs and clears the \nofillbit~bit of the MSHR with the same cache line address. 

\reffig{fig_timeline_authorize} shows two possible situations of a speculative load that has a cache miss and gets a no-fill MSHR. In the first case where the load is authorized early, its address can enter the \sht~before the requested cache line is returned. If the load address is selected for \shtfetch, the \shtnotify~signal will match the address of the no-fill MSHR. The \nofillbit~bit can be cleared as the address has been authorized and is safe to fill the cache. We show in \refsec{perf_rasspec_misses} that 46.48\% of speculative loads causing an L1D cache miss are authorized before the cache line is returned in benign benchmarks.

%This mechanism applies to both ongoing load and store operations because store operations also need to fetch the cache line before writing store data 
\reffig{fig_block_diagram} shows an example of clearing the \nofillbit~bit of the MSHR of address 0x3000 with \shtnotify. The \shtnotify~signal is also sent to L2-MSHR and clears the \nofillbit~bit if the L2 cache also has a matching MSHR entry. \shtnotify~is a low-cost operation that only checks the MSHR and does not look up the \textit{Tag/Data Storage}.

In \rasspec, a non-speculative load or store can also clear the \nofillbit~bit of a no-fill MSHR for the same cache line.%, which triggers a \shtnotify~sent to the next cache level.

%Another signal sent together with \shtfetch~is the \shtnotify~signal, which can improve the performance by enabling cache fills. The signal affects no-fill requests that have a cache miss but have not been returned. If the request is authorized and inserted into the \sht, the selected \shtfetch~address may match the MSHR entry. Upon an address match, the \shtnotify~signal can clear the \nofillbit~bit of the MSHR. An example is shown in \reffig{fig_block_diagram} where the MSHR for the address 0x3000 is cleared.

\iffalse

\bheading{\sht~operations.}
\sht~is a hardware unit that generates requests called \shtfetch~which are sent to the cache and allowed to fill in new cache lines. Different from defenses like InvisiSpec \cite{invisispec} and GhostMinion \cite{ghostminion} which store a local copy of data, \sht~only stores the address to generate \shtfetch.  

To mitigate speculative execution attacks, only addresses of memory accesses that have been resolved to be non-speculative can be inserted into \sht. A memory store is always sent to the cache after the corresponding instruction commits so we insert store addresses into \sht~when the request enters the cache. A memory load is considered non-speculative (\textbf{Authorized}) if all previous instructions have finished execution without a fault. This means the load cannot be a covert sending due to an unresolved prediction or address computation, an unauthorized access with permission denied, or a nullified instruction due to a faulting instruction, which are vulnerabilities exploited by speculative execution attacks. 

To identify authorized memory accesses, the re-order buffer (ROB) which tracks the active instructions in the pipeline is modified to mark each instruction as being speculative or not. When a memory access is authorized, its address can be inserted into \sht. 
%~and does not require coherence check for authorized accesses.

When an \shtfetch~enters the cache, it first checks if the address is already in the \textit{Tag/Data Storage} (see \reffig{fig_block_diagram}). If the address is not found, the \shtfetch~request is inserted into MSHR as a missing request. \shtfetch~always has the \nofillbit~bit cleared.
%The most straightforward idea is to store recently accessed memory accesses that are not allowed to fill the cache due to the \textit{\nofill} protection.

\bheading{Random Offset for \shtfetch}
While using authorized addresses for \shtfetch~prevents speculative execution attacks, fetching the original authorized address is vulnerable to side-channel attacks. In a non-speculative side-channel attack, %a victim program is authorized to access secrets and secret-dependent addresses so 
all addresses in the program will become \sht~entries. Issuing \shtfetch~to these addresses into the cache is equivalent to replaying the accesses and still leaks the secret.

We propose to take a random displacement (offset) from the selected \sht~address while enforcing the no-fill policies. \textit{Filling a different cache line than the demand fetch address decorrelates both cache reuses and evictions in side-channel attacks.} The required range (window) of random offsets is dicussed in \refsec{sec_analysis_window_size}.

\ifthesis{
\begin{figure}[t]
    \centering
    \includegraphics[width=\linewidth]{figures_ras/block_diagram_l1l2.pdf}
    \caption{The block diagram of L2 cache and its signals.}
    \label{fig_block_diagram_l1l2}
\end{figure}
}

\bheading{Constant-rate \shtfetch}
We propose to issue \shtfetch~at a constant rate. This can further prevent cache occupancy attacks \cite{cacheoccupancy} which measure the frequency of cache evictions during execution. When all the demand fetches are made no-fill and the cache will install new lines only for the \shtfetch, constant-rate \shtfetch es lead to a constant eviction rate, hence defeating the attack. In fact, as long as the issue rate does not depend on data used in execution, it can prevent cache occupancy attacks.

\bheading{Random Replacement}
Commonly used cache replacement policies such as LRU and PLRU are known to leak secrets \refsec{sec_bg_side_channel_ras}. We choose the random replacement policy to prevent such leakage. Random replacement policy is stateless, making it infeasible to record or recover secret-dependent state changes in either non-speculative or speculative channels.

\begin{figure}[t]
    \centering
    \ifelsethesis{
        \includegraphics[width=0.45\linewidth]{figures_ras/sht_queue_rand.pdf}
    }
    {
        \includegraphics[width=0.6\linewidth]{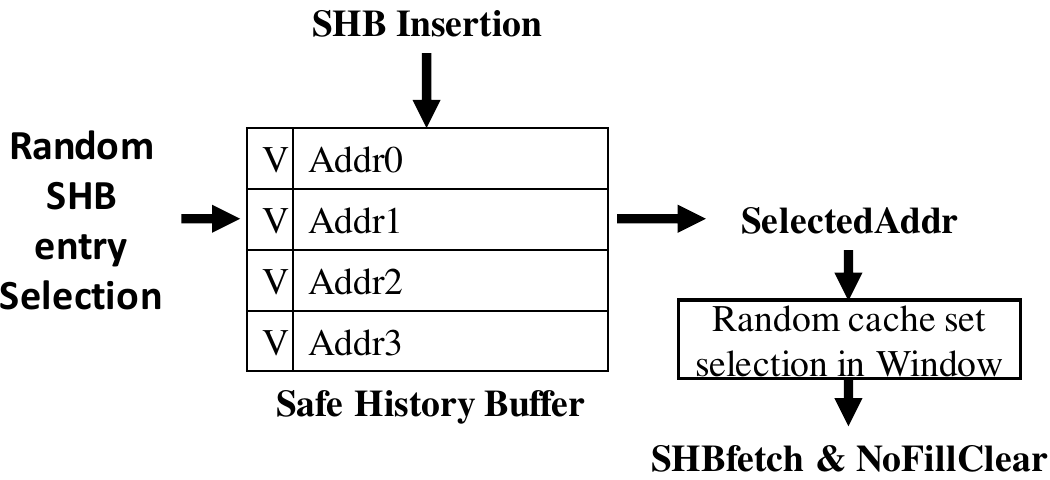}
    }
    \caption{The structure of a queue-type \sht. A random \sht~entry is selected for \shtfetch.}
    \label{fig_sht_queue_rand}
\end{figure}

\bheading{Design flexibility.} \sht~has multiple places for the designer to make design choices while providing the same level of security. We describe a few configurations.

\begin{packeditemize}
    \item \sht can be organized differently for insertion and lookup. \sht~can be a queue-like storage which saves recently inserted addresses. \sht~can also be a direct-mapped structure where an addresses can only take a certain entry.
    
    \item \sht~can have different selection policies for choosing an entry to generate \shtfetch. One option is to randomly select an entry. Other heuristics can also be used.% such as sending a fixed number of \shtfetch es for an entry before dropping this entry and move to the next one.

    \item When an address can be marked non-speculative (authorized) and inserted into \sht~can change according to the threat model, e.g., only checking unresolved branches.
\end{packeditemize}

Even after the design time, the software still has some flexibility to configure the defense, e.g., by writing to hardware control registers. We evaluate the performance of the following three parameters in \refsec{sec_perf_ras}

\begin{packeditemize}
    \item The issue rate of \shtfetch.

    \item While the \sht~hardware has a fixed number of \sht~entries (addresses), the software can control the number of active entries during runtime.

    \item The window size of random offsets added to generate \shtfetch.

    \item The window could be shifted in different ways by adjusting the lower bound.
\end{packeditemize}

\subsection{Preventing Cache Fills}
\label{sec_ras_nofill}

Cache fills are the essential cause of cache timing attacks. \ras~records whether a memory request is allowed to fill the cache by adding a new field called \nofillbit~(see \reffig{fig_block_diagram}) to the memory request signal, miss status holding registers (MSHRs) and writeback buffer entries. A no-fill memory request is handled normally if it has a cache hit. If a request has a cache miss, an MSHR is allocated and records the \nofillbit~status of the request.

When a memory request gets the requested cache line from the next level of memory, the cache line is placed in a buffer called the line fill buffer and the \nofillbit~bit of the corresponding MSHR is checked. Only loads and stores whose \nofillbit~bit is not set will fill the new cache line into the data and tag storage (the \textit{Fill} path in \reffig{fig_block_diagram}\ifthesis{ and \reffig{fig_block_diagram_l1l2}}). 

For a no-fill load, the cache will only forward the requested data through the \textit{NoFill} path to the processor (from the L1D cache in \reffig{fig_block_diagram}) or the upper-level (faster) cache\ifthesis{ (from the L2 cache in \reffig{fig_block_diagram_l1l2})}.

For a no-fill store, the store data is written to the line fill buffer. The cache line is then written back without filling the cache. The write-back of this dirty cache line is also marked no-fill so that the next-level cache will directly forward the line into its writeback buffer\ifthesis{ (see the \textit{No-fill Write} path in \reffig{fig_block_diagram_l1l2}}.

\bheading{Updating \nofillbit~Status by \shtnotify}
\ifelsethesis{\shtfetch~is asynchronous because whether an address of a memory access is used for \shtfetch~is independent of whether the memory access is complete. Specifically, if}{If} a memory instruction has a long cache miss and is authorized early, its address can enter \sht~before the requested cache line is returned. As \ras~fetches adjacent lines of \sht~addresses, it is possible that an \shtfetch~has the same address as one ongoing memory request which is no-fill.

We propose a technique called \shtnotify~to allow the cache fill in this case. A missing no-fill request is recorded in MSHR. If \sht~sends an \shtfetch~to the same address, the \nofillbit~status can be cleared. This mechanism applies to both ongoing load and store operations because store operations also need to fetch the cache line before writing store data 

\ifelsethesis{
\reffig{fig_block_diagram} shows that \sht~sends the same address for both \shtfetch~and \shtnotify. If there is no MSHR with the same address, \shtnotify~changes nothing and is dropped. If there is an MSHR allocated to a no-fill request with the same address as \shtnotify, the \nofillbit~bit can be cleared. When the cache line is returned, the cleared \nofillbit~bit will allow the line to fill the cache.

If a matching MSHR is found in L1-MSHR, the \shtnotify~signal further propagates to the L2 cache because the ongoing request may also have a miss in L2. \reffig{fig_block_diagram_l1l2} shows that \shtnotify~is sent to the MSHR in the L2 cache. L2-MSHR tracks the memory requests that have not been returned from the memory. \shtnotify~clears the \nofillbit~bit if there is a matching MSHR entry or otherwise gets dropped. \shtnotify~is a fast and low-cost operation that only checks the MSHR and does not incur the overhead of looking up the \textit{Tag/Data Storage} and delaying any other memory accesses.
}
{
\reffig{fig_block_diagram} shows that \sht~sends the same address 0x3000 for both \shtfetch~and \shtnotify. The \nofillbit~bit of the MSHR entry with the same address 0x3000 can be cleared. The \shtnotify~signal is further sent to L2-MSHR and clears the \nofillbit~bit if there is also a matching MSHR entry. \shtnotify~is a low-cost operation that only checks the MSHR and does not look up the \textit{Tag/Data Storage}.
}

Without \shtnotify, no-fill demand fetch accesses can prevent \shtfetch~requests from filling the corresponding cache sets, which compromises the security by causing evictions in random cache sets. Furthermore, \shtnotify~can benefit the performance by allowing more cache fills while preserving the security.

\fi
\section{Security Analysis}
\label{sec_analysis_window_size}

\rasspec~with an arbitrary window size can always defeat speculative cache channels as the cache fills by \shtfetch~are based on authorized addresses. Similarly, \rasall~never makes speculative cache state changes.

\rasall~decorrelates cache fills from the demand fetch by adding a random offset within a window. The window size affects \rasall 's security against non-speculative side-channel attacks. We first discuss the security impact of window size against contention-based and reuse-based attacks. We then discuss the security of \ras~against other cache-related attacks.

\subsection{Impact of Window Size.}
%The window size of \rasall~affects its security against non-speculative cache channels. 

\bheading{Contention-based attacks in L1 cache.} \rasall~can prevent the contention-based attacks in the set-associative L1 cache by making it equally likely for cache sets to be evicted and filled. To achieve maximum security by covering all cache sets, the window size can be set to a multiple of the number of cache sets multiplied by the cache line size, i.e., a multiple of the cache way size.
\ifthesis{This way, \ras~can achieve a randomized L1 cache using a conventional set-associative cache. 
It gives maximum entropy and uncertainty to the attacker as to which cache set the victim program used.}

This hardware-dependent window size has the advantage that \textit{the hardware can automatically set the window size for maximum entropy}. Software programmers do not have to figure out the security-critical regions to set the window size, and existing software can be protected without modification.

\bheading{Reuse-based attacks.}
We can also mitigate non-speculative reuse-based attacks if the window covers the security-critical region accessed with a secret-dependent index, meaning every cache line in the region has an equal chance of being fetched.

The cache collision attack tries to observe a shorter execution time when the victim will reuse certain secret-dependent addresses in the security-critical region. Having all such addresses equally likely to be fetched will eliminate the timing difference. The flush-reload attacker measures the time it takes to reload each cache line in the region. Noticing a cache hit at a random address in the region will still give no information. In both cases, the attacker cannot recover the secret value.

If the security-critical region is larger than the window size, random fetches within the window can protect the lower bits of an address covered by the window\footnote{For instance, if the window size is the number of cache sets in a way, i.e., the way size, the most significant bits (MSBs) of a coarse-grained attack using the way size as a unit, can leak information.} but cannot provide absolute security. This window size dependent security is labeled $\checkmark_W$ in \reftbl{tbl_matrix_defense_dimension}.
%This conditional protection against cache collision attacks has been the best security guarantee achieved by Random Fill Cache \cite{randomfill} ($\checkmark_W$ in \reftbl{tbl_matrix_defense_dimension}).

\ifthesis {
\bheading{Flexibility about window configuration.} Randomly fetching an address near a previously accessed address can improve performance if a program has a good spatial locality. While there is a requirement for window size for security, the window could be shifted in different ways by adjusting the lower bound. The lower bound can be set to the exact \sht~address for forward prefetching,  (\sht~address - $\frac{WindowSize}{2}$) for adjacent prefetching and other options. In our \ras~implementations, the lower bound is set to be the lower bound of the WindowSize-aligned region that contains the selected \sht~address (formally defined in the following equation).
\begin{equation}
    LowerBound = Addr-(Addr \mod WindowSize)
\end{equation}
}

\bheading{Security-performance trade-offs.}
A large window size may degrade the likelihood of a future cache hit on the selected line fetched since temporal and spatial locality may be lost across this wide window. Conceptually, a smaller window has better performance than a larger one as shown in \reffig{fig_perf_ras_plus}.

\subsection{Other Cache-related Attacks}

\ifthesis{
While reuse-based and contention-based attacks are the most critical cache timing attacks, we discuss the security and potential modifications to mitigate other related attacks.
}

\bheading{Speculative Interference Attacks.} In speculative interference attacks \cite{speculativeinterference}, speculative instructions change microarchitectural states and affect the timing of non-speculative instructions that are earlier in program order (older) but execute later than the speculative instructions. \ifthesis{For instance, if a speculative instruction executes earlier and keeps a specific functional unit busy, an older instruction of the same type can be delayed due to resource contention. The attacker essentially leverages backward-in-time interference and detects the speculative microarchitectural state change with the older instructions. }
%There are four types of speculative interference attacks exploiting the cache: two involving cache line evictions and fills and two involving use of the missing status handling register (MSHR) usage.
Speculative interference attacks can exploit cache evictions or fills.

To exploit speculative cache fills, the first attack evicts the cache line, which the older instruction will use, with the speculative cache line. This contention-type interference will increase the execution time of the older instruction. The other is speculatively fetching the cache line the older instruction will use. This reuse-type interference will shorten the execution time of the older instruction. \textbf{\ras~defenses already defend against these attacks because no speculative cache fill can happen.}

\ifelsethesis{As a part of the cache, the MSHR can cause similar attacks. If speculative accesses take all available MSHRs, the older instruction will be delayed due to MSHR contention. In another case, if a speculative load to the same address as the older memory instruction has a cache miss and gets an MSHR, the older memory instruction that executes later will reuse this MSHR and have a shorter access time. The simplest solution is to delay sending speculative memory requests until all previous memory instructions have their addresses resolved. An alternative is to use a GhostMinion \cite{ghostminion} type defense to allow the older instruction to take over an MSHR.
}
{There are similar attacks targeting the MSHR. A solution against the MSHR leakage is to allow the older instruction to take over an MSHR as done in GhostMinion \cite{ghostminion}.}

\bheading{Speculative cache coherence state change.} A speculative load can leak secret information by changing the exclusive state of a cache line in another private cache to a shared state. \ifthesis{In a MESI coherence system example, this can be a transition from a Modified (M) or Exclusive (E) state in a remote cache to a Shared(S) state. In a coherence-based speculative execution attack, the receiver can run on a remote core and prepare a cache line in the M state by writing to it. The sender running on another core requests the same address and changes the cache line to the S state. The receiver then performs another write to the address, and the execution time will be longer to invalidate the copy in the sender's cache before doing the write. }As a defense against this attack, the speculative modification from the M state to an S state should not be allowed, and the speculative load needs to re-execute when authorized.

\bheading{Occupancy-based attacks.} The attacker exploiting the cache occupancy channel primes the cache with his cache lines and measures the number of evicted cache lines due to the victim's behavior. The \rasall~defense can add noise to the attacker's observation by decoupling the number of evictions from the victim's accesses by allowing cache fills and evictions only for constant-rate \shtfetch es. However, \rasall~cannot completely prevent occupancy-based attacks. For instance, a victim may execute a long or short program depending on the secret value. Even with the constant-rate \shtfetch~enabled, there will be more cache evictions for the long program and fewer cache evictions for the short program. Further restrictions on the susceptible victim execution are needed against this attack.
%A counter-example is  the victim execution time is secret-dependent

\bheading{Prefetcher-based attacks.} Our \ras~defense is a safe prefetching approach. We do not combine it with other prefetchers. Detailed consideration of security of different prefetchers in combination with \ras~is beyond the scope of this paper and a possible topic for future work.

\section{Security Evaluation}
\label{sec_eval_security_ras}

We evaluate the security of \rasspec~and \rasall~with a suite of representative attacks. We use the GEM5 \cite{gem5} simulator to implement a baseline architecture without protection and defenses. The hardware parameters and defense functionalities are shown in \reftbl{tbl_hw_config_ras}. The parameters, e.g., \#\sht~entries and window size, are the settings that have the best performance in \refsec{sec_perf_ras}. 
%All the tested systems use the same set-associative caches commonly used in computer processors.
%\shtdmrand~and \shtfamri~both have two variants whose name has a suffix \earlyinsert~for early insertion of authorized addresses and \lateinsert~for late insertion of authorized addresses. Early and late insertion do not affect the security against cache timing attacks and are only compared for performance overhead in \refsec{sec_perf}.
We run \rasspec~against a contention-based and a reuse-based Spectre v1 attacks. We run \rasall~against the same two Spectre v1 attacks and four different types of cache side-channel attacks on the Advanced Encryption Standard (AES) encryption.%, covering critical columns in \reftbl{tbl_matrix_defense_dimension}.

\begin{table}[t]
    \centering
    \ifelsethesis{
        \includegraphics[width=0.65\linewidth]{figures_ras/hw_config.pdf}
    }
    {
        \includegraphics[width=0.95\linewidth]{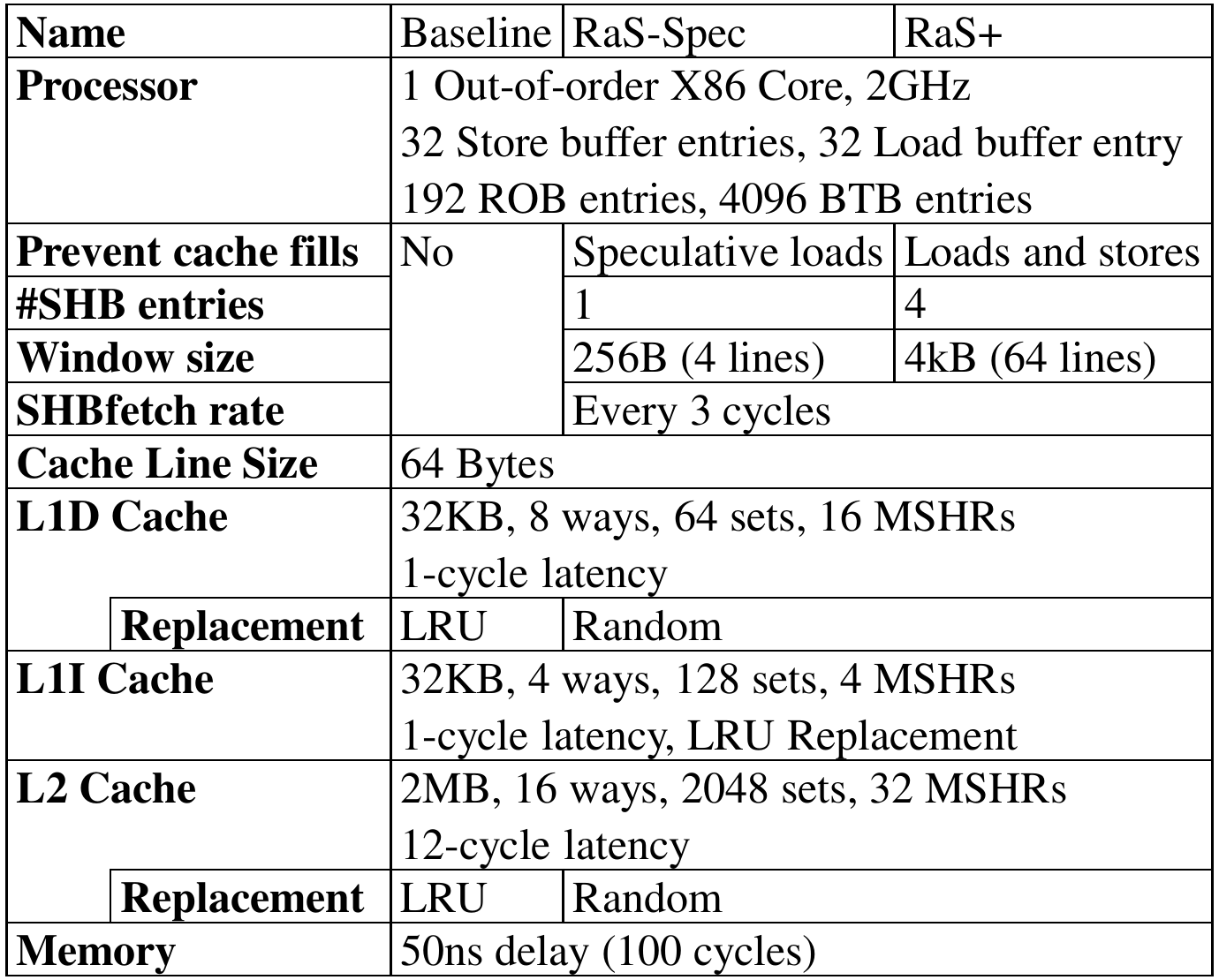}
    }
    \caption{\normalsize{Configurations of simulated systems in GEM5.}}
    \vspace{-20pt}
    \label{tbl_hw_config_ras}
\end{table}

\ifthesis{
\begin{figure}[t]
    \centering
    \ifelsethesis{
        \includegraphics[width=0.6\linewidth]{figures_ras/eval_aes_pp.pdf}
    }
    {
        \includegraphics[width=0.85\linewidth]{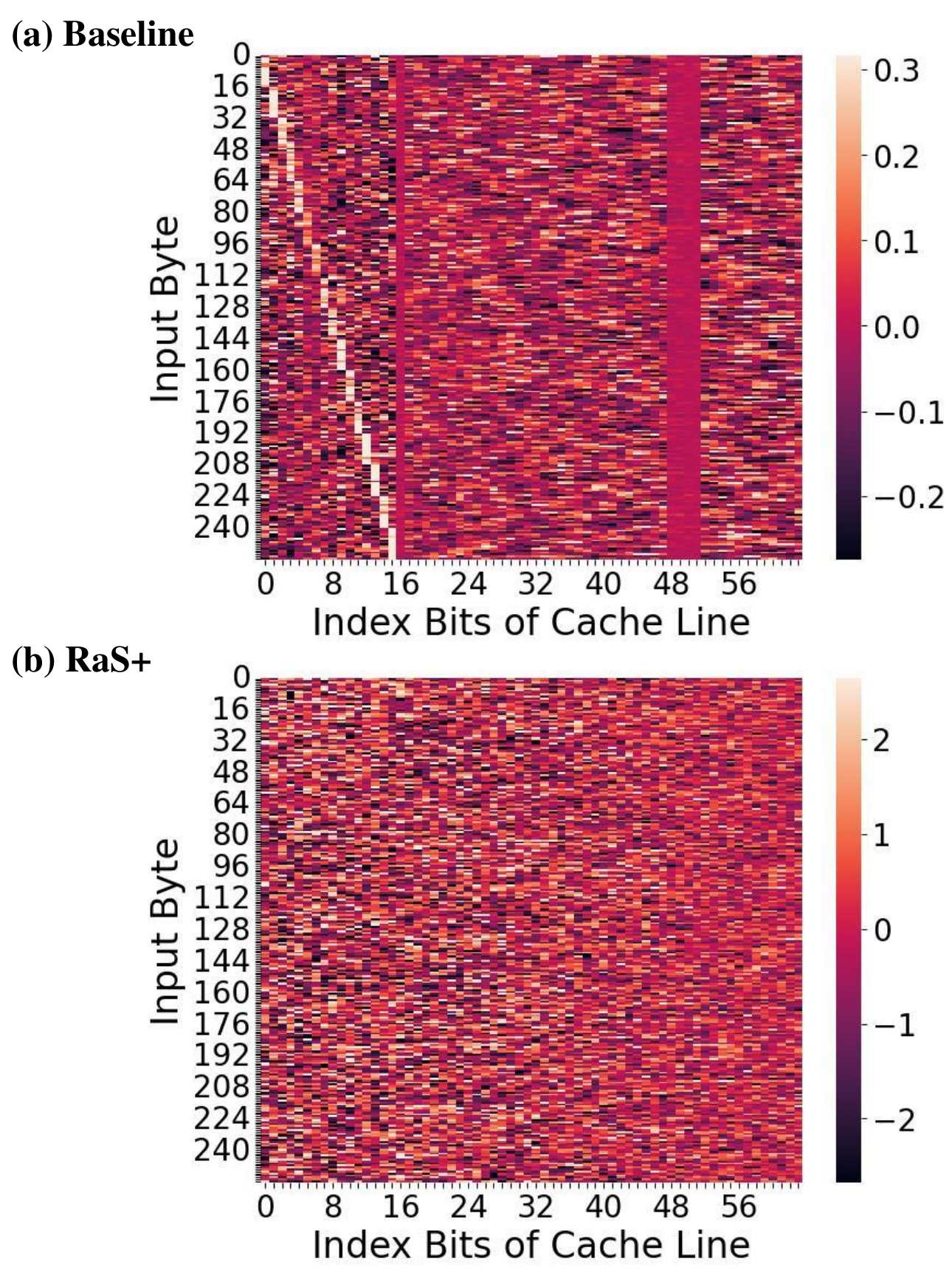}
    }
    \caption{The attacker's measurement in a prime-probe side-channel attack on AES. The key byte is 0. A lighter pixel stands for a longer access time.}
    \vspace{-10pt}
    \label{fig_eval_aes_pp}
\end{figure}

\begin{figure}[t]
    \centering
    \ifelsethesis{
        \includegraphics[width=0.6\linewidth]{figures_ras/eval_aes_fr.pdf}
    }
    {
        \includegraphics[width=0.85\linewidth]{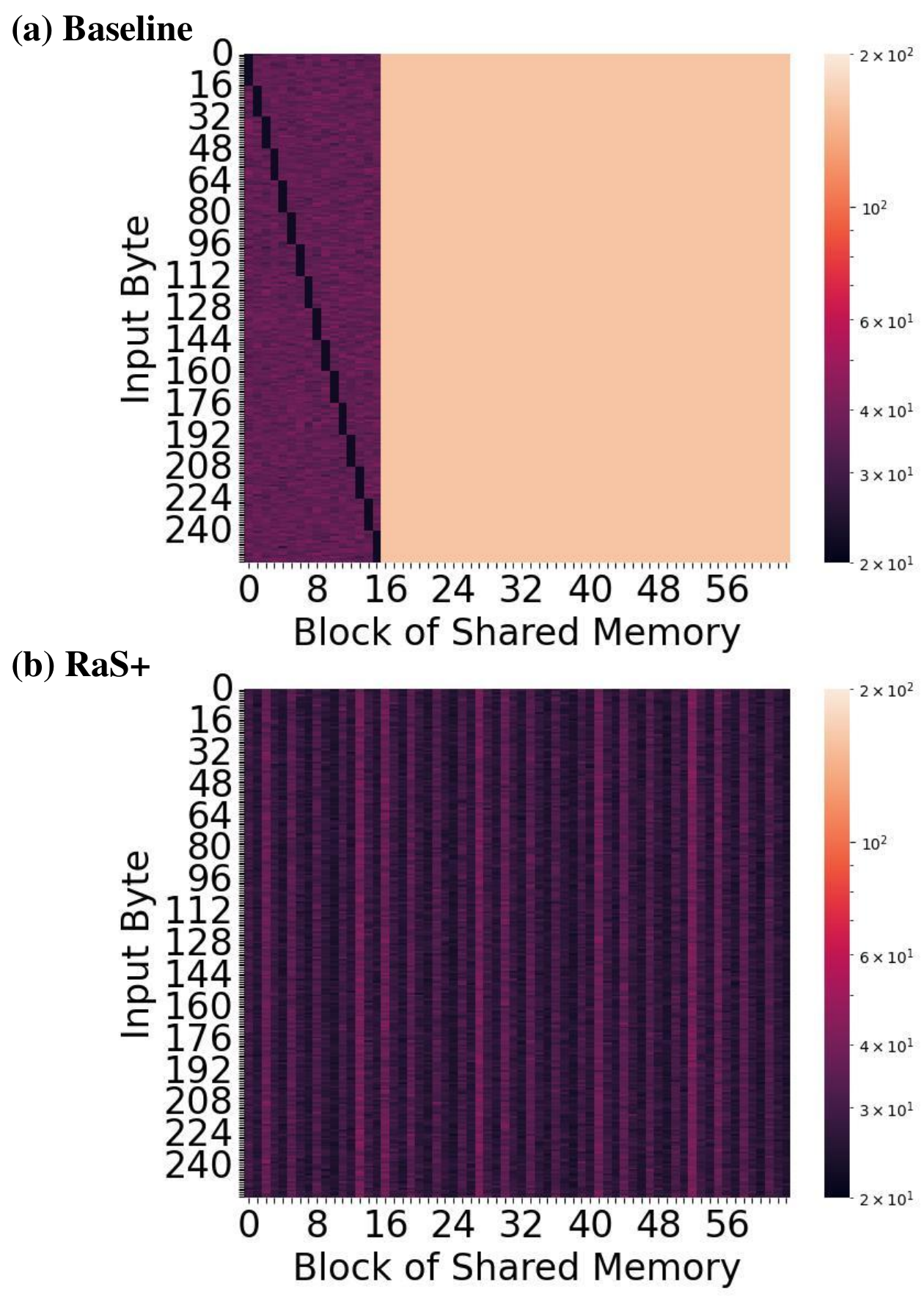}
    }
    \caption{The attacker's measurement in a flush-reload side-channel attack on AES. The key byte is 0. A darker pixel stands for a shorter access time.}
    \label{fig_eval_aes_fr}
\end{figure}
}

\subsection{Spectre v1 Attack}

Spectre v1 attack mistrains the predictor for conditional branches to access an out-of-bounds secret speculatively. We run Spectre v1 attacks, which use either a flush-reload or a prime-probe cache channel. \ifthesis{In the attacks, the code emulates a strong attacker who does the same-domain measurement of cache states.}

\ifthesis{
\begin{figure}[t]
    \centering
    \includegraphics[width=0.6\linewidth]{figures_ras/code_spectre_v1.pdf}
    \caption{\normalsize{The key operations of a flush-reload Spectre-v1 Attack. 
    \ifthesis{
    Mistraining the hardware predictor and flushing all cache lines of the \textit{shared} array happen before the sender code and are not shown. Commonly used values of \textit{step}, the step size of flushing and reloading, are 64, 512 or 4096. However, the attacker can use a different value if he controls the sender.}}}
    \label{fig_code_spectre_v1_ras}
\end{figure}
}

\bheading{Flush-reload cache channel.} \ifthesis{\reffig{fig_code_spectre_v1_ras} shows the key operations of a flush-reload attack. }\reffig{fig_eval_spectre_v1} (a) to (c) shows the attacker's measurement with or without the defense to recover the secret value of 30.
\ifthesis{In the Baseline system, the sender code can successfully fill the cache line of \textit{shared[30*64]} into the cache, causing a shorter access time to that address in the reloading phase. The secret is recovered as 30.}
In \rasspec~and \rasall~systems, the speculative access to \textit{shared[30*64]} cannot fill the cache.
\ifthesis{As the execution never commits, the address of \textit{shared[30*64]} will not be inserted into \sht~and later fetched.}
%We test bigger values, e.g., 4096, for \textit{step}. The result also shows a successful attack in the Baseline system while the \ras~and \rasplus~systems defeat the attack.

\bheading{Prime-probe cache channel.} \ifthesis{In the speculative execution, the sender code can also cause a secret-dependent eviction. }The attacker code tries to fill the cache with his cache lines before the speculative execution (Prime). The sender code accesses its array, e.g., array\_sender[secret$\times$64], to evict one attacker's line in a secret-dependent cache set. During the Probe phase, the attacker accesses his array and detects a longer access time for the evicted cache set, thus leaking the secret.

\begin{figure}[t]
    \centering
    \ifelsethesis{
        \includegraphics[width=\linewidth]{figures_ras/eval_spectre_v1.pdf}
    }
    {
        \includegraphics[width=0.95\linewidth]{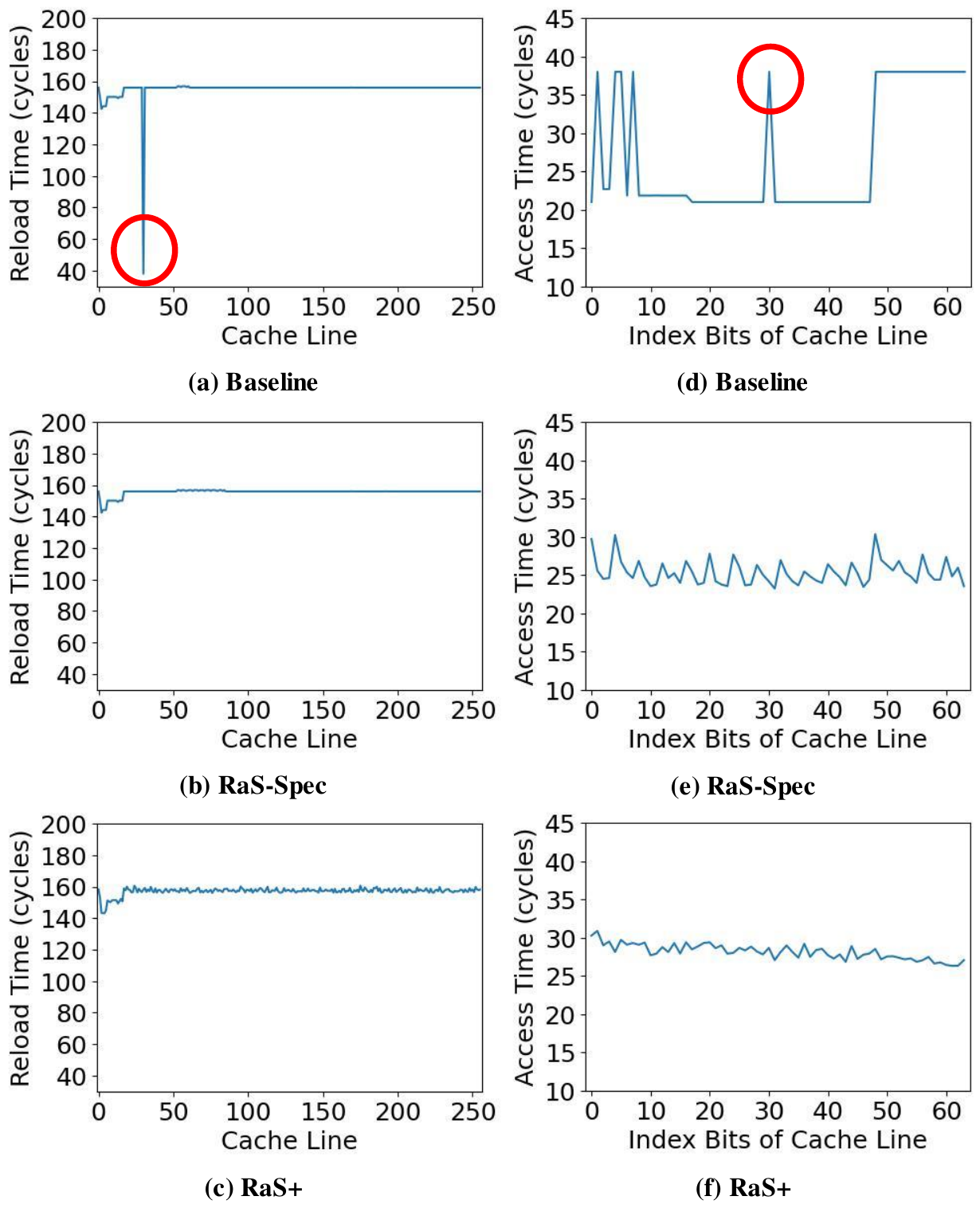}
    }
    \caption{\normalsize{(a) to (c): The attacker's measurement in a flush-reload Spectre v1 attack. (d) to (e): The attacker's measurement in a prime-probe Spectre v1 attack. The secret value is 30.}}
    \label{fig_eval_spectre_v1}
    \vspace{-10pt}
\end{figure}

\reffig{fig_eval_spectre_v1} (d) shows the secret-dependent eviction, which moves when the secret value changes, happens in the cache set 30 (circled)\ifthesis{, causing the secret value of 30 to be leaked}. Some other sets also have high access times caused by unrelated memory accesses. 
\ifthesis{The high access times for these cache sets do not change when a different secret value is used. These times are a common noise in contention-based cache attacks.}
The secret-dependent eviction no longer appears in \rasspec~(e) and \rasall~(f) systems. \ifthesis{This is because the defenses disallow the cache fill of speculative accesses so that the unauthorized execution will not evict any cache line. }

\ifelsethesis{}
{
\begin{figure*}[t]
    \centering
    \ifelsethesis{
        \includegraphics[width=\linewidth]{figures_ras/eval_aes_all.pdf}
    }
    {
        \includegraphics[width=0.98\linewidth]{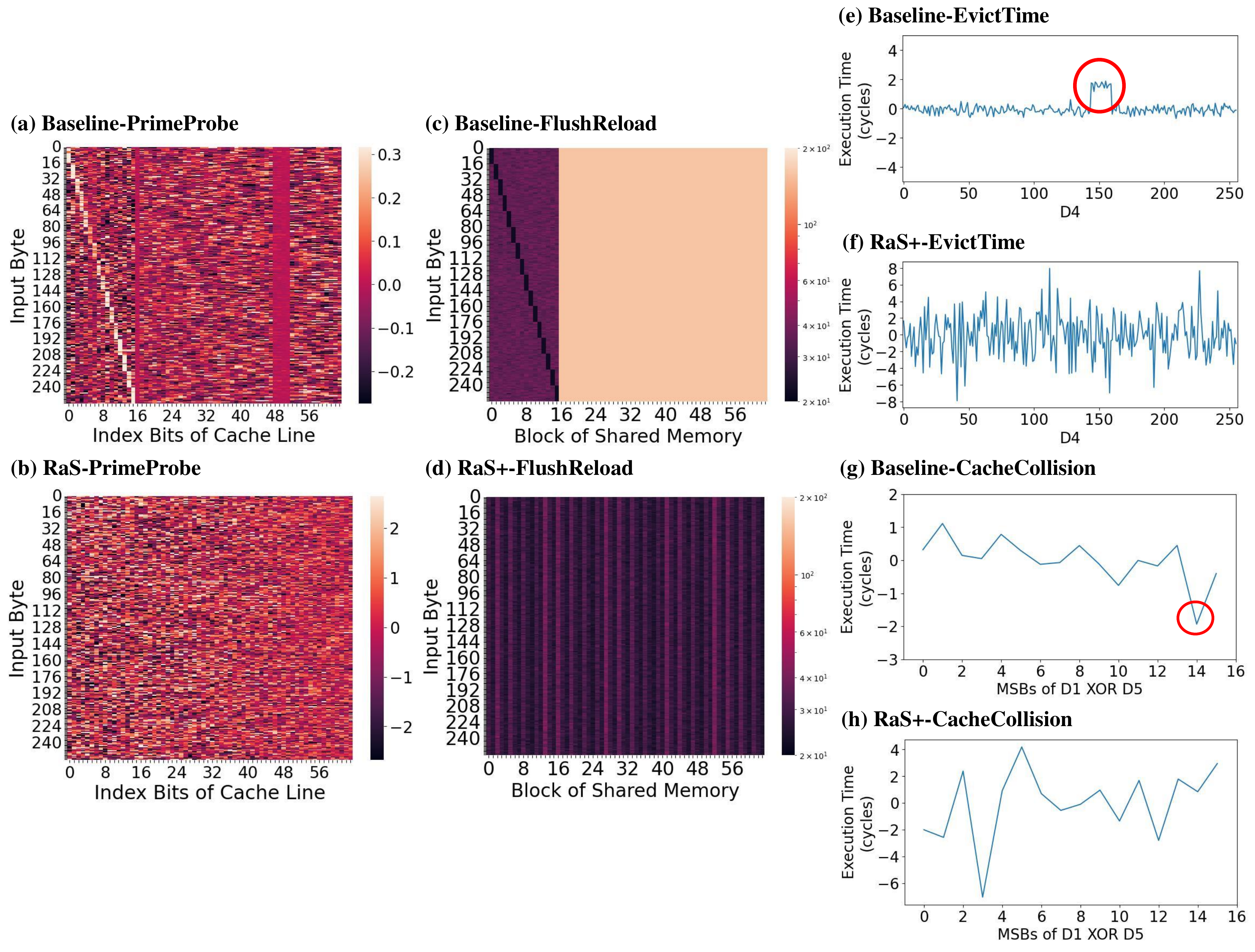}
    }
    \caption{\normalsize{The attacker's measurement when tested side-channel attacks run in the Baseline and \rasall.}}
    \label{fig_eval_aes_all}
    \vspace{-10pt}
\end{figure*}
}

\subsection{Cache-side channel attack on AES}
\label{sec_sht_eval_aes}

\ifthesis{
The Advanced Encryption Standard (AES) algorithm is a commonly used standard for data encryption and decryption. The secret key generates an address to read the AES lookup tables. If an attacker can observe the cache hits or misses caused by a victim's AES encryption program, the victim's key can be leaked.}
We test four access-based and operation-based cache side-channel attacks: prime-probe, flush-reload, evict-time and cache collision attacks. All four attacks exploit the first round of an AES-128 encryption program, which reads 4 AES lookup tables, $T_1$ to $T_4$,
%. Each table has 256 4-byte entries. The entries accessed in the first round are 
with the addresses shown below. $D_i$ and $K_i$ are 16 input bytes and key bytes. $\oplus$ is the XOR operation of two bytes.

$T_1[D_1 \oplus K_1], T_1[D_5 \oplus K_5], T_1[D_9 \oplus K_9], T_1[D_{13} \oplus K_{13}]$

$T_2[D_2 \oplus K_2], T_2[D_6 \oplus K_6], T_2[D_{10} \oplus K_{10}], T_2[D_{14} \oplus K_{14}]$

$T_3[D_3 \oplus K_3], T_3[D_7 \oplus K_7], T_3[D_{11} \oplus K_{11}], T_3[D_{15} \oplus K_{15}]$

$T_4[D_4 \oplus K_4], T_4[D_8 \oplus K_8], T_4[D_{12} \oplus K_{12}], T_4[D_{16} \oplus K_{16}]$

\ifthesis{
We assume a strong attacker who controls the input data and can accurately capture the victim's execution to do measurements. %The attacker measures the time of access or execution and computes the average time after the encryption is run for at least $2^{15}$ times with random input. 
%We show the effectiveness of \ras~defense when the victim enables the defense to protect the encryption program. 
The window size is chosen to be 4kB, which is a multiple of the 4kB cache way size and also large enough to cover the AES tables (see the window size requirement in \refsec{sec_analysis_window_size} to address both reuse-based and contention-based attacks). %This means \rasall~will fetch one of the 64 cache lines in the adjacent region of the selected \sht~address.
}

\ifthesis{
\begin{figure}[t]
    \centering
    \ifelsethesis{
        \includegraphics[width=0.56\linewidth]{figures_ras/eval_aes_et_twobytes.pdf}
    }
    {
        \includegraphics[width=0.9\linewidth]{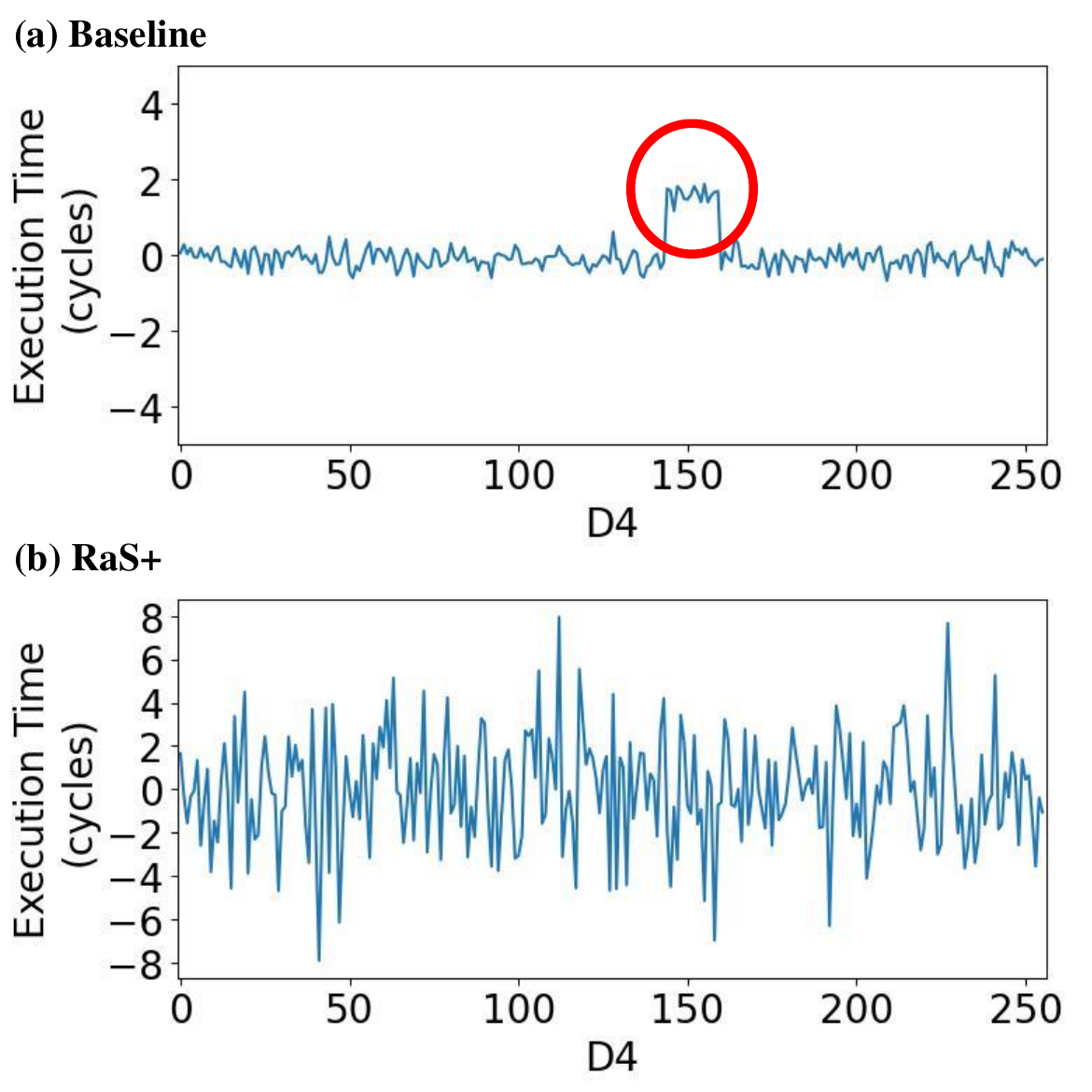}
    }
    \caption{The execution time measured by the attacker in an evict-time side-channel attack on AES. The average execution time is subtracted from the measured time to show the variance of time. X axis is the value of an input byte.}
    \label{fig_eval_aes_et}
\end{figure}

\begin{figure}[t]
    \centering
    \ifelsethesis{
        \includegraphics[width=0.6\linewidth]{figures_ras/eval_aes_cc.pdf}
    }
    {
        \includegraphics[width=0.9\linewidth]{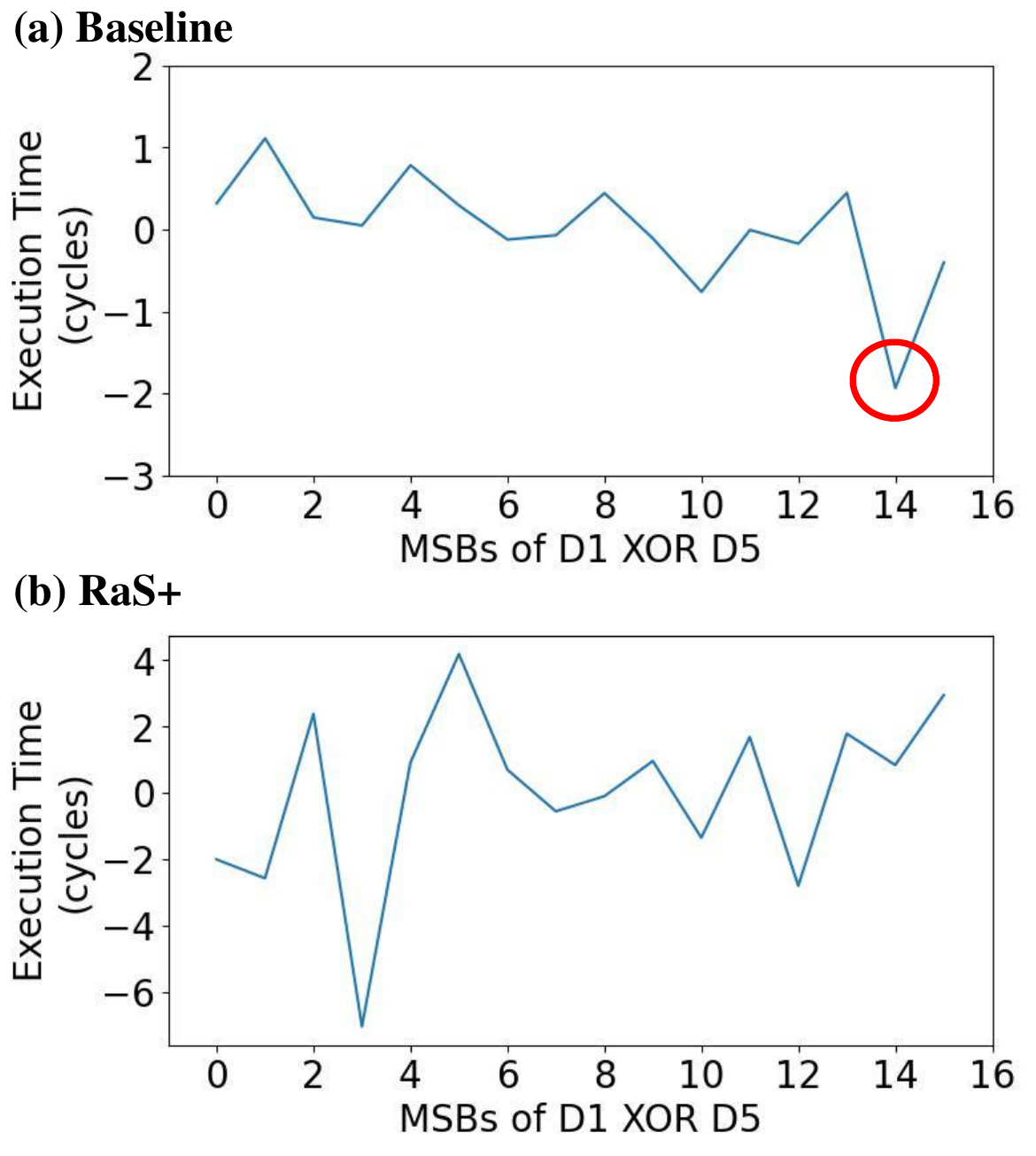}
    }
    \caption{The execution time measured by the attacker in a cache collision side-channel attack on AES. The average execution time is subtracted from the measured time to show the variance of time. X axis is the xor result of input bytes D1 and D5.}
    \label{fig_eval_aes_cc}
\end{figure}
}

\bheading{Prime-probe attack.} \ifthesis{A prime-probe attack on the AES encryption does not require shared memory. The attacker fills the cache by reading a big array before the victim's execution and measures the latency to access the addresses that will be mapped to different cache sets.

}
\ifelsethesis{\reffig{fig_eval_aes_pp} (a)}{\reffig{fig_eval_aes_all} (a)} shows the attacker's measurement in the Baseline system. The light diagonal contains the cache sets that have a higher access latency due to the contention with AES table accesses when different input values are used. The attacker can only recover the 4 MSBs of each key byte, which are also found to be 0x0. The four LSBs of the key byte cannot be determined, as accessing any of the 16 4-byte AES entries in the same cache line can cause the cache line to be fetched.
\ifelsethesis{\reffig{fig_eval_aes_pp} (b)}{\reffig{fig_eval_aes_all} (b)} shows that \rasall~can defeat the attack and no meaningful pattern can be found. 
\ifthesis{As the window size of 4kB is also the way size of the cache, each cache set has an equal chance to be accessed by the \shtfetch~and cause an eviction. The defense defeats the prime-probe attack.}

\bheading{Flush-reload attack.} \ifthesis{A flush-reload attack on an AES program is possible when the AES tables are in a memory region shared by multiple processes, e.g., as a part of a shared library. The attacker measures which cache lines of AES table entries are in the cache, which means the victim uses them. 

}
\ifelsethesis{\reffig{fig_eval_aes_fr} (a)}{\reffig{fig_eval_aes_all} (c)} shows that one key byte can be partially leaked in the Baseline system. The attacker measures a 4kB region, i.e., 64 cache lines, where the first 1kB is the AES table $T_1$. With the y-axis being the input byte and the x-axis being the accessed line, the dark diagonal represents the XOR result of the input byte and the key byte. The 4 MSBs of this key byte can be found to be 0x0. 
\ifthesis{Similar attacks can be repeated for the other 15 key bytes, causing 4 bits in each key byte to be leaked.}
\ifelsethesis{\reffig{fig_eval_aes_fr} (b)}{\reffig{fig_eval_aes_all} (d)} shows the attacker's measurement with the \rasall~defense. 
\rasall~performs random fetching within the 4kB shared memory region, 
\ifelsethesis{which is covered by the window, so every cache line has an equal chance to be fetched. The first 16 lines of the AES table and the other 48 unused lines have similar access latency.}
{causing no key-dependent dark dots.
}

\ifelsethesis{}
{
\begin{figure*}[t]
    \centering
    \includegraphics[width=0.9\linewidth]{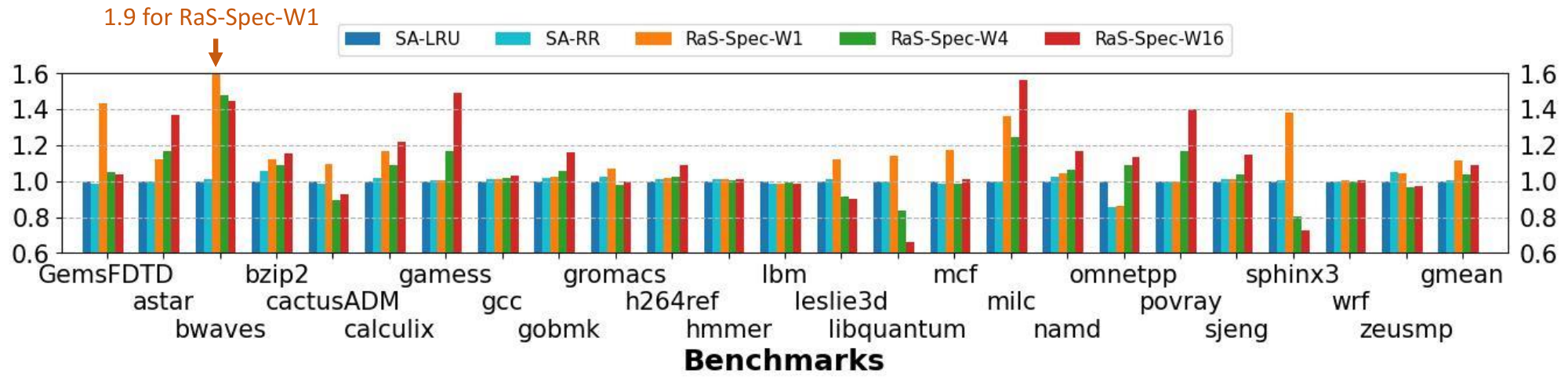}
    \caption{\normalsize{Normalized execution time in the SA-LRU Baseline, \sarr, and \rasspec~defenses. The last set is the geometric mean of normalized execution time. The high execution time of bwaves with \rasspec-W1 is labeled.}}
    \label{fig_perf_ras_spec}
    \vspace{-10pt}
\end{figure*}
}

\bheading{Evict-time attack.} \ifthesis{The example evict-time attack emulates a victim whose code evicts a specific cache set before the AES encryption. If the encryption uses the AES table entries mapped to the evicted set, the execution time will be longer.

}
\ifelsethesis{\reffig{fig_eval_aes_et} \ifelsethesis{(a) and (b) show}{(a) shows}}{\reffig{fig_eval_aes_all} (e)} the attacker's observation when a cache set containing some entries in the lookup table $T_4$ is evicted before encryption.
\ifelsethesis{Therefore, when the input bytes D4 and D8 are certain values, accessing $T_4[D_4 \oplus K_4]$ and $T_4[D_8 \oplus K_8]$ will take longer time, also increasing the total execution time. 

From \reffig{fig_eval_aes_et} (a) and (b), as a fixed cache set is evicted, the attacker can know $D_4 \oplus K_4 = D_8 \oplus K_8$ for $D_4$ and $D_8$ with longer execution time. Observing that the long-time D4 values are 144 to 159 (0x9* in hex) and long-time D8 values are 0xa* in hex, the attacker can infer that $K_4 \oplus K_8 = 0x3*$, which is the correct guess when $K_4$ is 0x65 and $K_8$ is 0x5e.}
{
Observing that the long-time D4 values are 0x9* in hex, the attacker can recover the four most significant bits (MSBs) of the key byte.
}
\ifelsethesis{\reffig{fig_eval_aes_et} \ifelsethesis{(c) and (d) show}{(b)}}{\reffig{fig_eval_aes_all} (f)} shows no meaningful longer execution time which can leak the key because the memory accesses in the eviction phase are made no-fill and unable to evict the target set.

\bheading{Cache-collision attack.} The reuse-based cache collision attack \cite{cachecollision} on AES observes a shorter execution time when certain input values cause a cache line containing AES table entries to be reused. We emulate a stronger attacker using 1 MSHR entry in the L1D cache instead of 16 MSHR entries in \reftbl{tbl_hw_config_ras}. Having fewer MSHR entries reduces the number of accesses other than the target access to cause cache collision, giving the attacker a clearer observation of low execution time.

\ifelsethesis{
\reffig{fig_eval_aes_cc} (a) shows a lower execution time due to reuse when accessing the lookup table $T_1$. When $T_1[D_1 \oplus K_1]$ and $T_1[D_5 \oplus K_5]$ access the same cache line, i.e., the MSBs of $D_1 \oplus D_5$ are equal to the MSBs of $K_1 \oplus K_5$, the execution time is shorter. The attacker can learn $K_1 \oplus K_5 = 0xe*$ from the figure, which is the correct guess as $K_1$ is 0x0f and $k_5$ is 0xe6.
}
{
\ifelsethesis{\reffig{fig_eval_aes_cc} (a)}{\reffig{fig_eval_aes_all} (g)} shows a lower execution time when $T_1[D_1 \oplus K_1]$ and $T_1[D_5 \oplus K_5]$ access the same cache line. $K_1 \oplus K_5$ is correctly revealed as 0xe*.
}
\ifelsethesis{\reffig{fig_eval_aes_cc} (b)}{\reffig{fig_eval_aes_all} (h)} shows a low execution time at a wrong location. The attacker will have a wrong guess of the key.
%This is because the fetching of AES entries is randomized by \ras, defeating the attack.

\section{Performance Evaluation}
\label{sec_perf_ras}

First, we show the impact of randomization in cache design. Next, we show the superior performance of \rasspec~compared to other hardware defenses against speculative execution attacks. For this purpose, we use the same GEM5 parameters (\reftbl{tbl_hw_config_ras}), simulation methodology and SPEC 2006 benchmarks as used by important defenses like InvisiSpec \cite{invisispec:correction}, CleanupSpec \cite{cleanupspec} and MuonTrap \cite{muontrap}.
Finally, we show the performance of \rasall~compared to secure caches to defeat cache side-channel attacks, like Random Fill Cache \cite{randomfill}.

\iffalse
We evaluate the performance impact of deploying \rasspec~and \rasall~defenses. As shown in \reftbl{tbl_hw_config_ras}, we use similar cache parameters as previous speculative execution defenses like Invisispec \cite{invisispec:correction} and Cleanupspec\cite{cleanupspec}, and we also use  24 SPEC CPU 2006 benchmarks.
% (UNCLEAR, DELETE) Note that since encryption-based randomized LLCs \cite{ceaser, scattercache} are set-associative caches, the L2 configuration can also model the timing with LLC protection.
%We measure the performance overhead to run 24 SPEC CPU 2006 benchmarks with the \ras~defenses. 
Each benchmark uses the reference data set. The first 10 billion instructions are skipped. The execution time of running the next 500 million instructions is measured. \textbf{Choosing the SA-LRU cache as the Baseline, skipping a fixed number of instructions and measuring the time for running the same number of instructions have been the common practice for evaluating hardware defenses \cite{invisispec:correction, cleanupspec, muontrap, ghostminion}, which we adopt to get comparable performance results for \rasspec.}

\ifthesis{For both systems, we study the impact of three parameters including the \shtfetch~rate, the number of \sht~entries and the window size. The setting with the best performance is shown in \reftbl{tbl_hw_config_ras} and used for security evaluation.

For sensitivity study, we change one parameter while fixing the others. Each defense is labelled with its parameter values. For instance, \rasall-R3E4W64 is a \rasall~defense which issues an \shtfetch~every 3 cycles, has 4 \sht~entries and uses a window size of 64 cache lines (4096 bytes).}

\fi

\iffalse
\begin{table}
    \centering
    \includegraphics[width=0.75\linewidth]{figures/perf_aes.pdf}
    \caption{\normalsize{Relative execution time of different numbers of AES runs in the Baseline and two \rasplus~systems.}}
    \vspace{-20pt}
    \label{tbl_perf_aes}
\end{table}

\subsection{AES Encryption}

We measure the performance overhead of running AES encryption in the protection mode of \rasplus~systems. The AES encryption is run for different numbers of times to encrypt a series of data. Decryption can also be used.

\reftbl{tbl_perf_aes} shows the relative execution time in the \shtdmrandpm~and \shtfamripm~systems. For a single AES run to encrypt one data block, \rasplus~systems have the overhead of 14\% and 40\% as the accesses to AES tables do not fill the cache and the \sht~does not have entries that can prefetch to-be-used data. As the number of AES runs increase to encrypt more blocks, addresses in the AES table are inserted into the \sht, which can trigger \shtfetch~requests to fetch adjacent addresses during execution. The relative execution time of \rasplus~systems gets closer to the Baseline system for a larger number of runs. \shtfamripm~even achieves 1\% improvement in performance when the number of AES runs is 100 thousand.
\fi

\subsection{Impact of Random Replacement}

We compare set-associative caches with LRU (SA-LRU) with set-associative caches with random replacement (\sarr) for both the L1D and L2 caches (see \reffig{fig_perf_ras_spec}). It turns out the average overhead is negligible at 0.3\%. At worst, the performance of bzip2 benchmark is 5.8\% worse. In the best case, the execution time of the omnetpp benchmark doing network simulation is reduced by 14.4\%.

\ras~also has random fetching in a window and random selection of authorized address sites from the \sht, whose impact we show in the following sections.
\subsection{\rasspec~Performance} 
\label{sec_perf_rasspec}

\ras~architectures can be tuned with three design parameters: issue rate, number of \sht~entries and window size (denoted Rx, Ey and Wz). For example, \ras~with R5E4W16 means a prefetch rate of an \shtfetch~every 5 cycles, a \sht~with 4 entry and a window size of 16 cache lines.

We ran experiments with \sht~issue rates of one \shtfetch~per 3, 5, 7, and 10 cycles and found that one \shtfetch~per 3 cycles gives the best performance for almost all benchmarks. The cache lines prefetched by \shtfetch~are useful for later memory accesses, and fetching more lines leads to lower performance overhead.

We also ran experiments with different numbers of \sht~entries. Having only one entry (E1), i.e., keeping only the most recently authorized address for \shtfetch, performs best for 22 out of 24 benchmarks.

\begin{table}[t]
\centering
\includegraphics[width=0.9\linewidth]{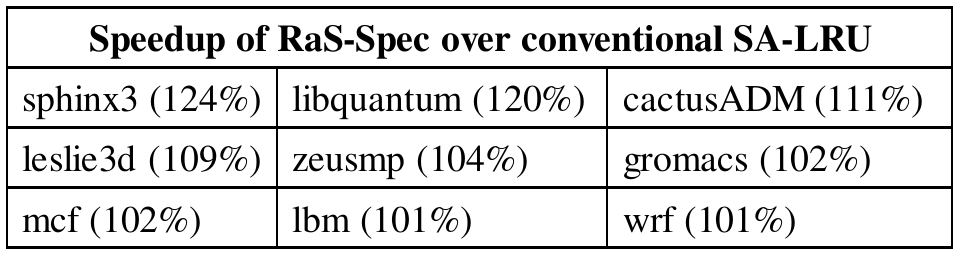}
\caption{\normalsize{Benchmarks which have speedup with \rasspec-W4 over the insecure Baseline.}}
\vspace{-20pt}
\label{tbl_perf_imprv_ras}
\end{table}

We experimented with window sizes of 1 (W1), 4 (W4) and 16 (W16) cache lines (see \reffig{fig_perf_ras_spec}).
W1 has a 11.4\% average slowdown while W16 has a 8.6\% average slowdown. The best window size is W4, which has only a 3.8\% average slowdown. The result shows that \textit{fetching randomly selected locations near an authorized address (example: W4) can improve the performance over fetching only the authorized address (W1)}.

\begin{figure*}[t]
    \centering
    \includegraphics[width=0.9\linewidth]{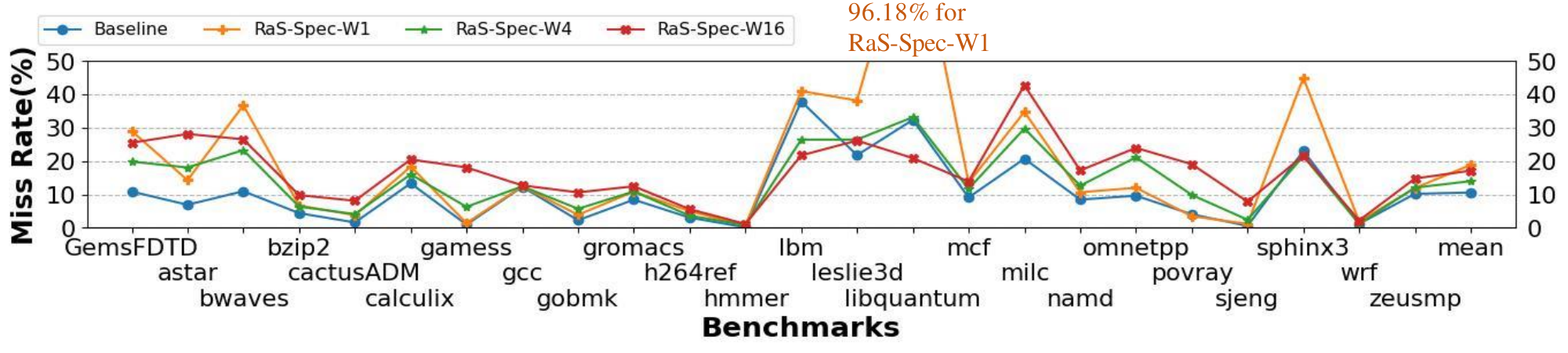}
    \caption{\normalsize{Cache miss rates in the Baseline and \rasspec~defenses with different window sizes. \textit{Mean} at the end is the average miss rate. The high miss rate of libquantum with \rasspec-W1 is labeled (outside the figure).}}
    \label{fig_perf_missrate_rasspec_l1d}
    \vspace{-10pt}
\end{figure*}

The best average performance of \rasspec~is 3.8\% for \rasspec-R3E1W4, i.e., the defense with the issue rate of one \shtfetch~per 3 cycles, 1 \sht~entry and a window size of 4. In fact, \reftbl{tbl_perf_imprv_ras} shows that it improves the performance of 9 benchmarks. 

We compare \rasspec~with important previous defenses with a similarly comprehensive threat model for speculative execution attacks, including misprediction, store-to-load forwarding and faults. InvisiSpec-Future \cite{invisispec:correction} has an overhead of 16.8\%. CleanupSpec \cite{cleanupspec}, an undo-type defense, has 5.1\% overhead. MuonTrap \cite{muontrap} has 4\% overhead. Our \rasspec~has a lower overhead of 3.8\%. Only GhostMinion \cite{ghostminion} reports a lower overhead of 2.5\%. Furthermore, \rasspec~only requires a small \shtfull~(1 entry) while InvisiSpec, MuonTrap and GhostMinion require a separate speculative cache structure. CleanupSpec causes new attacks \cite{unxpec}.

\subsection{Deeper Look into \rasspec~Performance}
\label{perf_rasspec_misses}

We now try to explain the performance benefits of \rasspec. \reffig{fig_perf_missrate_rasspec_l1d} shows the cache miss rates of L1D cache of the Baseline and the \rasspec~defenses with window sizes of 1, 4 and 16. The average cache miss rate of the Baseline is 10.57\%. The lowest miss rate of \rasspec~defenses is 13.98\% with a window size of 4 (W4). W16 has an average miss rate of 17.13\%. The worst miss rate of 18.87\% is for W1.

A window size of 1 can be disastrous for the cache miss rates of some benchmarks like libquantum (96.18\%), sphinx (45.00\%) and bwaves (36.73\%). The miss rate of the Baseline is 37.90\% for lbm, which is significantly improved with W4 (26.44\%) and W16 (21.72\%). The results show that \textit{the random fetching of a cache line within a window (W4, W16) can improve the performance (over W1)}.

\bheading{No-fill Optimization.} The no-fill feature prevents speculative execution attacks by not changing the cache state on cache misses. Unfortunately, it is also the main cause of performance degradation. We reduce the performance overhead by turning No-fill to Fill for cache misses whenever possible.

\begin{table}[t]
    \centering
        \includegraphics[width=\linewidth]{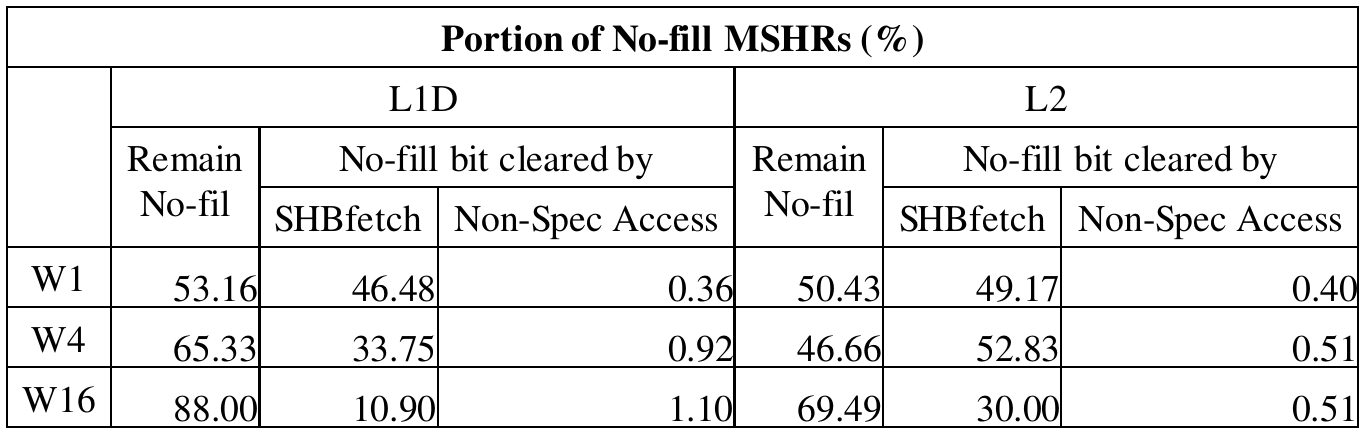}
    \caption{\normalsize{Average percentages of no-fill MSHRs in \rasspec~defenses whose no-fill bit is never cleared or cleared by a later \shtfetch~or non-speculative access.}}
    \label{tbl_perf_split_rasspec_mshr_tbl_avg}
    \vspace{-20pt}
\end{table}

\rasspec-W1 will only send the original address of a speculative load that gets authorized. The portion of MSHRs cleared by \shtfetch~(46.48\% in \reftbl{tbl_perf_split_rasspec_mshr_tbl_avg}) shows that \textit{nearly half of the speculative loads causing an L1D cache miss can be authorized before the cache line is returned so that the NoFill can be turned into a Fill and allowed to fill the cache.}

The percentages of MSHRs cleared by \shtfetch~in W4 (33.75\%) and W16 (10.90\%) are lower than W1 because, with a larger window, it is less likely for the original authorized address to be selected for \shtfetch, which clears the \nofillbit.

In \reftbl{tbl_perf_split_rasspec_mshr_tbl_avg}, the percentage of L2 MSHRs cleared by \shtfetch~is always higher than L1D MSHRs because the L2 miss latency, i.e., the time taken to access the memory, is high. Hence, it is more likely for a speculative access to be authorized before the requested cache line is returned. The high portion shows that \textit{the expensive L2 misses benefit more from \shtnotify~compared to L1D misses}.

We also measured the performance of \rasspec~without our optimization to clear \nofillbit~bits. The overheads of W1, W4 and W16 were 20.7\%, 17.1\% and 17.6\%, respectively, but we have reduced them to 11.4\%, 3.8\% and 8.6\% in \refsec{sec_perf_rasspec} with our no-fill optimization.
\begin{figure*}[t]
    \centering
    \includegraphics[width=\linewidth]{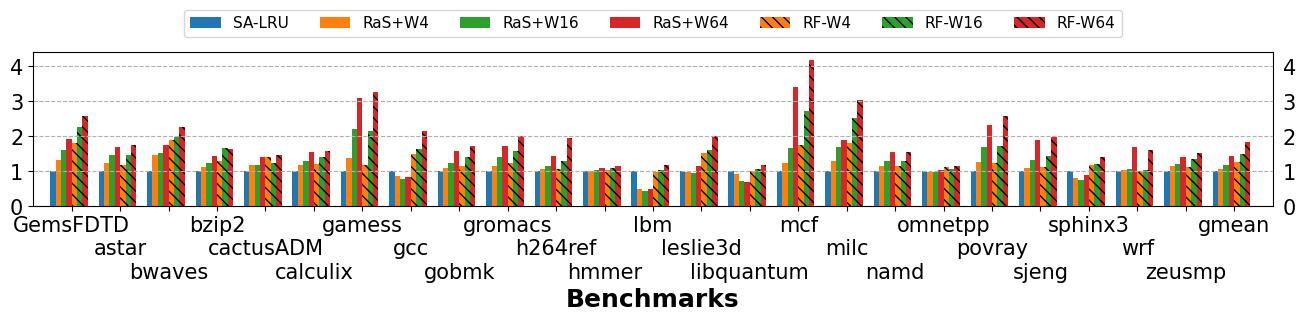}
    \caption{\normalsize{Normalized execution time of \rasall~and Random Fill cache (RF) protecting both L1D and L2 caches.
    The same color is used for defenses with the same window size. 
    The last set is the geometric mean of normalized execution time.}}
    \label{fig_perf_ras_plus}
\end{figure*}

\subsection{\rasall~Performance}
\label{sec_perf_rasall}

We want to compare \rasall~architectures with defenses against non-speculative cache timing attacks.

% may move to later
For the strongest protection against contention-based attacks, we want \shtfetch~to randomly choose among all cache sets, of which there are 64 sets in L1D cache. This is equivalent to the random selection of a cache line in a window of 64 cache lines.
For best security protection, we used a large window of W64 for our initial experiments on the issue rate and \sht~entries below.

%Like \rasspec, 
We ran experiments with issue rates of one \shtfetch~per 3, 5, 7, and 10 cycles and found that one \shtfetch~per 3 cycles has the best performance for all benchmarks.

We tested \sht~with 1, 4 and 16 entries (denoted E1, E4 and E16) while fixing R3 and W64. The interesting finding is that E4 has a lower average performance overhead (45.2\%) than E1 (56.0\%) and E16 (47.7\%).

The performance improvement with E4 can be due to the quality of addresses in the \sht. A high-quality address is the address whose adjacent lines are used in later execution. \sht~with one entry can easily lose a high-quality address when the next authorized address is inserted. \sht~with more entries can keep a high-quality address longer for prefetching. However, if there are too many entries in the \sht, e.g., 16 or even more entries, old addresses will reduce the chance of using newer addresses for \shtfetch, leading to less useful prefetching. E4 is the option that preserves the quality and freshness of addresses for most benchmarks.

\begin{table}[t]
\centering
\includegraphics[width=0.9\linewidth]{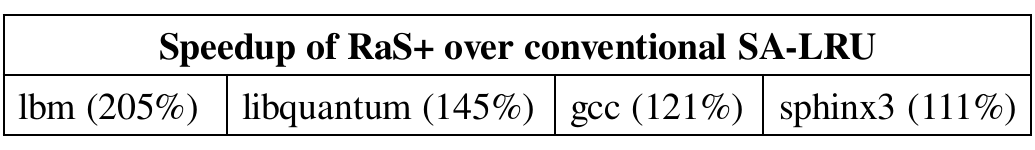}
\caption{\normalsize{Benchmarks which have speedup with \rasall W64 over the insecure Baseline.}}
\label{tbl_perf_imprv_rasplus}
\vspace{-20pt}
\end{table}

\reffig{fig_perf_ras_plus} shows the overhead of \rasall~with window sizes of 4, 16 and 64 cache lines (denoted W4, W16 and W64).
The average overheads of W4, W16 and W64 compared to the Baseline are 7.9\%, 18.8\% and 45.2\%, respectively, as shown by the geometric means (gmean of \rasall~bars in \reffig{fig_perf_ras_plus}).

The high performance overhead of W64 prompted us to consider security-performance trade-offswith smaller window sizes. Interestingly, our security tests showed the the cache collision side-channel attacks is still defeated with smaller window sizes in \rasall W16 and \rasall W4.
%Future work can provide new security metrics that quantify different levels of security protection provided.

The low overhead of W4 makes it a defense option to always turn on to add small disturbance to an attacker's observation. W64 has a large average slowdown but it does provide guaranteed security against all attacks leaking the cache set number.
Surprisingly, \rasall W64 can improve the performance of some benchmarks significantly (see \reftbl{tbl_perf_imprv_rasplus}).

These results show that \rasall~allows security-performance trade-offs: larger window sizes provide more security guarantees but lower performance while smaller window sizes offer higher performance.

\subsection{Deeper Look into \rasall~Performance}
\label{perf_rasplus_misses}

We found that cache miss rates of \rasall~defenses with window sizes of 4, 16 and 64 cache lines are higher than the SA-LRU Baseline by 4.59\%, 9.10\% and 21.88\%, respectively. This shows that the average cache miss rate increases with the window size.

\begin{table}[t]
    \centering
        \includegraphics[width=\linewidth]{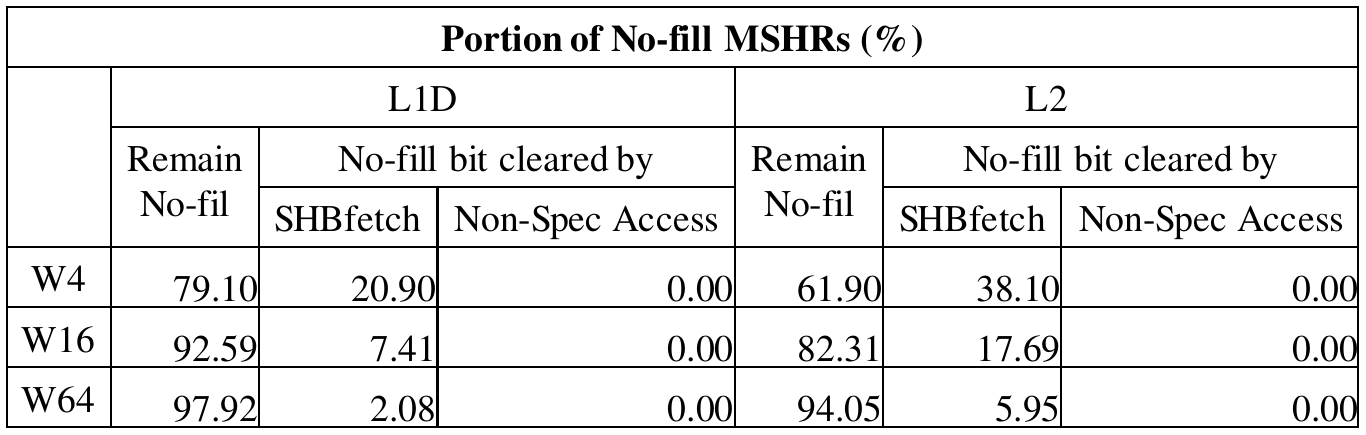}
    \caption{\normalsize{Average percentages of no-fill MSHRs in \rasall~defenses whose no-fill bit is not cleared or gets cleared by a later \shtfetch.}}
    \label{tbl_perf_split_rasplus_mshr_tbl_avg}
    \vspace{-20pt}
\end{table}

\bheading{No-fill Optimization.}
\reftbl{tbl_perf_split_rasplus_mshr_tbl_avg} shows the percentages of MSHR entries where NoFill can be turned to Fill in \rasall. Unlike \rasspec, all loads and stores are labeled no-fill in \rasall~and cannot clear the \nofillbit~bit of previous accesses (zeros for columns of Non-Spec Access in \reftbl{tbl_perf_split_rasplus_mshr_tbl_avg}).

Similar to \rasspec, the percentage of MSHRs with NoFill's cleared to Fill's decreases with a larger window size. Also, no-fill bits of L2 misses in \rasall~are more likely to be cleared by the no-fill optimization than L1D misses.

We also measured the performance of \rasall~without our optimization to clear \nofillbit~bits. The overheads of W4, W16 and W64 were 42.3\%, 35.3\% and 57.8\%, respectively, but we have reduced them to 7.9\%, 18.8\% and 45.2\% in \refsec{sec_perf_rasall} with our no-fill optimization.
\subsection{Performance Comparison with Side-channel Defenses}
\label{sec_perf_rfill}

%We compare \rasall~with the Random Fill (RF) cache architecture \cite{randomfill}, which defeats non-speculative reuse-based attacks and has been the only solution against the most challenging cache collision attack. We show that \rasall, which defeats more attacks, performs better than RF cache.

Since \rasall~defeats non-speculative cache timing attacks, not just speculative execution attacks, we want to compare it with hardware defenses for non-speculative cache timing attacks. For cache timing attacks (\refsec{sec_bg_side_channel_ras}), the most challenging is the cache collision attacks. So far, the only defense published for this is the Random Fill (RF) cache \cite{randomfill}. We show that \rasall, which defeats many more attacks, performs better than RF cache.

\ifthesis{
\begin{figure*}
    \centering
    \includegraphics[width=\linewidth]{ch-ras/figures_ras/perf_ras_cmp_rfill.jpg}
    \caption{Performance comparison of \rasall~and Random Fill (RF) cache architecture.}
    \label{fig_perf_ras_cmp_rfill}
\end{figure*}
}

Unlike \rasall, RF cache protected only the L1 cache but not other cache levels like the L2 cache. RF cache may generate a prefetch request based on a speculative address, which can lead to successful speculative execution attacks, whereas \rasall~generates prefetch requests only based on non-speculative address sites in \sht. Also, RF cache generates a prefetch request for each cache miss whereas \rasall~issues an \shtfetch~at a constant rate, e.g., every 3 cycles, to decorrelate when cache fills happen from when cache misses happen.

For comparison, we implemented RF cache with protection for both L1D and L2 caches, as we did for \rasall~caches. \reffig{fig_perf_ras_plus} compares the performance of \rasall~and the RF cache for three window sizes each. RF-W4, RF-W16 and RF-W64 have average performance overheads of 26.7\%, 49.3\% and 84.5\% respectively.
\rasall W4, \rasall W16 and \rasall W64 have performance overheads of 7.9\%, 18.8\% and 45.2\%, which perform better than RF cache at each window size.

%For each window size, \rasall~has better performance than RF cache by 18.8\%, 30.5\% and 39.3\% for W4, W16 and W64, respectively. \rasall~outperforms RF in most of the benchmarks.
%In fact, \rasall-W4 outperforms RF-W4 in 19 out of 24 benchmarks. \rasall-W16 and \rasall-W64 outperform RF-W16 and RF-W64 in 22 benchmarks.

\section{Comparison with Past Works}

\subsection{\rasspec~vs Previous Speculative Defenses}
\label{sec_ras_vs_past_spec_def}

\bheading{1) Delay-based speculative defenses} such as NDA \cite{specshield} and STT \cite{stt} delay the execution of speculative loads. \rasspec~does not block the execution of loads, which is critical to provide load data to dependent instructions.

\bheading{2) Redo-type speculative defenses} such as InvisiSpec and GhostMinion \cite{ghostminion} require a second access for the authorized addresses. Different from these defenses, \rasspec~does not always require a second access. \rasspec~has a novel quick authorization feature, \shtnotify, that only needs to clear the no-fill status if the access is quickly authorized so that no extra access is needed. We find that \shtnotify~can clear the no-fill bit for 46.48\% of speculative loads in \rasspec-W1. \shtnotify~reduces the performance overhead from 20.7\% to 11.4\% for RaS-Spec-W1 and from 17.1\% to 3.8\% for RaS-Spec-W4.
%of \rasspec-W1, \rasspec-W4 and \rasspec-W16 by 9.3\% (20.7\% to 11.4\%), 13.3\% (17.1\% to 3.8\%) and 9.0\% (17.6\% to 8.6\%), respectively.
%In addition, \rasspec~has a buffer of addresses instead of a buffer of data that needs write-backs and validation for coherence.

\bheading{3) Undo-type speculative defenses} such as CleanupSpec \cite{cleanupspec} restores the cache state if the execution is squashed. Restoration of cache state is complicated and introduces the new unXpec attack \cite{unxpec}.

As a result, \rasspec~has an average performance overhead of only 3.8\%, the best performing defense against speculative execution attacks except for GhostMinion which claims a 2.5\% overhead.

\subsection{\rasall~vs Combinations of Previous Defenses}

We emphasize that it is not sufficient to just naively combine a defense for speculative execution attacks with a different defense for non-speculative cache timing attacks. Some mechanisms of one defense can negate the protections of the other defense. Subtle new attack can be introduced. Often, the hardware required is more than what is needed and the performance overhead is greater.

%Previous defenses are proposed to address different attacks. However, it is non-trivial to provide mitigate speculative and non-speculative attacks with a simple combination and avoid causing new attacks, which we do with \rasall.

First, not all defenses are compatible for combination. The undo-type speculative defense, CleanupSpec \cite{cleanupspec}, always allows cache fills. CleanupSpec is incompatible with the Random Fill Cache \cite{randomfill}, which disallows all demand-fetch cache fills.

Second, simply combining defenses can lead to new attacks. For instance, a combination of a redo-type speculative defense, InvisiSpec \cite{invisispec}, and a side-channel defense, Random Fill Cache \cite{randomfill}, is expected to defeat both speculative and non-speculative attacks. However, Random Fill Cache allows sending prefetch requests based on a speculative address. The attacker can reload the adjacent region of a speculative access to detect the prefetch, nullifying the protection by InvisiSpec to disallow cache state changes for the speculative access.

Third, combining defenses can lead to unnecessary performance overhead. For instance, a combination of a delay-based speculative defense, STT \cite{stt}, and a side-channel defense, Random Fill Cache \cite{randomfill}, always has worse performance than having only the Random Fill Cache due to extra delays of speculative accesses. \rasall~improves the performance upon Random Fill Cache (see \refsec{sec_perf_rfill}) while defeating both speculative and non-speculative cache timing attacks.

Fourth, there are attacks uncovered by previous defenses. For instance, the new writeback attack (see \refsec{sec_ras_nofill}) is not considered by previous speculative defenses that focus on only speculative loads. The attack is not considered by Random Fill Cache which focuses on only the L1 cache.

\section{Conclusions}

We introduce a new class of Random and Safe (RaS) cache architectures that defeat a wide range of cache timing attacks. RaS generates safe and useful addresses for fetching and filling memory lines into the cache, without leaking cache states that an attacker can correlate with the victim's memory accesses. \ras~defenses update no-fill status whenever possible to allow secure cache fills and improve the performance. RaS-Spec provides security from speculative attacks with a low performance overhead of 3.8\% without adding separate data storage. RaS+ can also protect against non-speculative side channels with different security-performance trade-offs. For some benchmarks, RaS can improve both performance and security over conventional insecure caches. RaS has a big design space that allows the designer to explore other implementations.

%%%%%%% -- PAPER CONTENT ENDS -- %%%%%%%%

%%%%%%%%% -- BIB STYLE AND FILE -- %%%%%%%%
\clearpage
\bibliographystyle{IEEEtranS}
\bibliography{refs}
%%%%%%%%%%%%%%%%%%%%%%%%%%%%%%%%%%%%

\end{document}